\documentclass[preprint,showpacs,preprintnumbers,amsmath,amssymb]{revtex4}

\usepackage{graphicx}% Include figure files

\newcommand{\og}[0]{~\rq\rq}
\newcommand{\fg}[0]{\lq\lq~}

\newcommand{\ket}[1]{\left|#1\right\rangle}
\newcommand{\bra}[1]{\left\langle#1\right|}

\def\*#1{\mathbf{#1}}

\begin{document}

\title{Giant Optical Non-linearity induced by a Single Two-Level
System interacting with a Cavity in the Purcell Regime}% Force line breaks with \\

\author{Alexia Auff\`eves-Garnier$^{1}$}%
%\altaffiliation[Also at ]{ECLyon}%Lines break automatically or can be forced with \\
\author{Christoph Simon$^{1}$}%
\author{Jean-Michel G\'erard$^{2}$}%
\author{Jean-Philippe Poizat$^{1}$}%
\affiliation{$^{1}$CEA/CNRS/UJF Joint team " Nanophysics and
semiconductors ",Laboratoire de Spectrom\'etrie Physique (CNRS
UMR5588),Universit\'e J. Fourier Grenoble1,140 rue de la Physique,
BP 87, 38 402 Saint-Martin d'H\`eres C\'edex, France}

\affiliation{$^{2}$CEA/CNRS/UJF Joint team " Nanophysics and
semiconductors", CEA/DRFMC/SP2M, 17 rue des Martyrs,38054 Grenoble,
France}
%This line break forced with \textbackslash\textbackslash}%
\email{alexia.auffeves-garnier@ujf-grenoble.fr}

\date{\today}% It is always \today, today,

\begin{abstract}
A two-level system that is coupled to a high-finesse cavity in the
Purcell regime exhibits a giant optical non-linearity due to the
saturation of the two-level system at very low intensities, of the
order of one photon per lifetime. We perform a detailed analysis of
this effect, taking into account the most important practical
imperfections. Our conclusion is that an experimental demonstration
of the giant non-linearity is feasible using semiconductor
micropillar cavities containing a single quantum dot in resonance
with the cavity mode.
\end{abstract}

\maketitle

\section{Introduction}

The implementation of giant optical non-linearities is of interest
both from the fundamental point of view of realizing strong
photon-photon interactions, and because it is hoped that such an
implementation would lead to applications in classical and quantum
information processing. One particularly promising system for
realizing large non-linearities is a single two-level system
embedded in a high-finesse cavity, which serves to enhance the
interaction between the emitter and the electromagnetic field. In
the so-called strong coupling regime, where the interaction between
the emitter and the light dominates over all other processes
including cavity decay, there are well-known dramatic non-linear
effects such as normal-mode splitting \cite{thompson}, vacuum Rabi
oscillations \cite{brunerabi} and photon blockade \cite{birnbaum}.

State of the art technology allows the realization of high-quality
semiconductor quantum dots and optical microcavities. A single
quantum dot at low temperature can be considered to a large extent
as an artificial atom, and can be manipulated coherently as a
two-level system under resonant excitation of its fundamental
optical transition. In particular, Rabi oscillations have been
observed between the first two energy levels of a quantum
dot~\cite{rabi}, and coherent operations on these two levels have
been realized~\cite{controlecoh}. Many quantum optics experiments
first realized with atoms become possible, including cavity quantum
electrodynamics experiments and the generation of quantum states of
light. While there have been several pioneering experiments for
semiconductor microcavities containing single quantum
dots~\cite{semicon}, the conditions for strong coupling are quite
challenging. On the contrary, the so-called Purcell
regime~\cite{purcell46,gerard99}, where the interaction between the
emitter and the cavity mode dominates over that with all other
modes, but where the cavity decay is still faster than the emitter
lifetime, is significantly easier to attain. In particular, it has
been reached for single-photon sources based on micropillars
containing quantum dots~\cite{Solomon,Moreau01,Var05}. It is
therefore of interest to consider the potential for large optical
non-linearities in the Purcell
regime~\cite{turchette95A,hofmann03,wakspra,waksprl}.

A pioneering experiment on optical non-linearities in the Purcell
regime was performed with atoms in a free-space cavity in a slightly
off-resonant configuration \cite{turchette95A}. The theoretical
study realized in Ref. \cite{hofmann03}, based on
the``one-dimensional atom'' model suggested in Ref.
\cite{turchette95B}, shows that for the case of a one-sided cavity
and for exact resonance between the light and the emitter, the
non-linearity is enhanced. This is due to the very simplest
non-linear effect, namely those related to the saturation of a
single two-level system by light that is in, or close to resonance
with the two-level transition. The coupling between the light and
the dipole is governed by the intensity of the light. When the
intensity is sufficiently high, the dipole becomes saturated and
thus effectively decouples from the light. Since the saturation
occurs at intensity levels of order one photon per lifetime of the
emitter, this effectively realizes a strong interaction between
individual photons, that is to say, a giant optical non-linearity.
This result has been the starting point of our work.

In the present work we study the potential of a quantum dot
interacting in the Purcell regime with a semiconducting microcavity
to realize a giant optical non-linearity. We have two main
motivations. First, we aim at deriving the quantum coupled mode
equations describing the dynamics of a two-level system placed in a
high finesse cavity, based on input-output theory developed in Ref.
\cite{Gardiner85}. Coupled mode equations indeed are often used by
semiconductor physicists and it seemed interesting to us to derive
them in the quantum frame in a rigorous manner.  This allowed us to
generalize the results of Ref. \cite{hofmann03} to non-resonant
situations and to double-sided cavities. The generalization to
multi-ports cavities is interesting in the perspective to exploit
the giant non-linearity in more complex architectures like add-drop
filters~\cite{akahane05}. Besides, we have included leaks and
excitonic dephasing in the model, which was mandatory as we wanted
to study the non-linear effect using realistic experimental
parameters. To our knowledge, this is the first extensive study of
this optical system including leaks and dephasing in the linear and
non-linear regime.

Our second motivation is to use the theoretical model to study the
feasibility of an experimental demonstration of the non-linearity
with a semiconductor micropillar cavity containing a single quantum
dot. The results obtained in this study are very promising, since
striking optical features like dipole induced reflection or giant
non-linear behavior are observable with uncharged quantum dots and
state of the art micropillars.

The paper is organized as follows. In section II, we establish the
coupled-mode equations for the cavity mode and for the input and
output fields. In section III the stationary solution of these
equations is derived in two regimes: first, we show that in the
linear case (low intensity excitation) the two-level system induces
a dip in the transmission of the optical medium. Second, we treat
the case of general intensities via a semi-classical approximation,
which allows to show the giant optical non-linearity. We devote
section IV to the generalization of the study to the case of leaky
atoms and cavities. In section V we discuss the relevance of the
two-level model to the case of a quantum dot and we use the model
developed in section IV to give detailed quantitative estimates of
the experimental signals we aim at evidencing. In particular, we
show that the non-linear effect is observable using state-of-the-art
microcavities.

\section{Quantum coupled-mode equations}

\begin{figure}[h,t]
\begin{center}
\includegraphics[height=5cm]{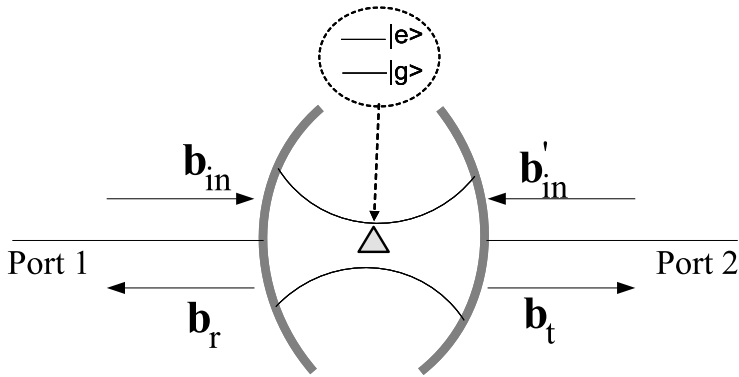}
\caption{\it Scheme of the atom-cavity coupled system. The atomic
frequency is $\omega_0$, the cavity mode frequency
$\omega_0+\delta$. The cavity mode is coupled to the outside world
via two ports labelled $1$ and $2$ modes with coupling constants
$g_1$ and $g_2$, the two-level system to the cavity mode with
coupling constant $\Omega$. The situation can describe a micropillar
containing a single quantum dot.} \label{fig:transmission}
\end{center}
\end{figure}

The situation considered is represented in
figure~\ref{fig:transmission}. A single mode of the electromagnetic
field is coupled to the outside world via two ports labelled $1$ et
$2$. Each port supports a one-dimensional continuum of modes
respectively labelled by the subscripts $_k$ and $_l$. This may
correspond to the case of a high finesse Fabry-Perot made of two
partially reflecting mirrors. Among the infinity of modes supported
by the cavity, we consider only one mode that interacts with two
continua of planewaves through the left and the right mirror. The
cavity contains a single two-level system of frequency $\omega_0$
which is nearly on resonance with the mode of interest. We note $a$,
$b_k$, $c_l$ the annihilation operator for the cavity mode, the
modes of port 1 and port 2 respectively, $\omega_0+\delta$,
$\omega_k$ and $\omega_l$ the corresponding frequencies. The atomic
operators are $S_z={\displaystyle
\frac{1}{2}(\ket{e}\bra{e}-\ket{g}\bra{g}})$ and
$S_-=\ket{g}\bra{e}$. The coupling strengths between the cavity and
the modes of port $1$ and $2$ are taken constant, real and equal to
$g_1$ and $g_2$ respectively. The total Hamiltonian of the system is
then

\begin{equation}
\begin{array}{l}
H=\hbar \omega_0 S_z + \hbar (\omega_{0}+\delta) a^+a + \sum_k \hbar
\omega_k b^+_k b_k + \sum_l \hbar \omega_l c^+_l c_l   +  \\
i\hbar \Omega (S_+a-a^+ S_-)+ \hbar \sum_k  (g_1 b_k^+ a-g_1 a^+ b_k
) + \hbar \sum_l (g_2 c_l^+ a- g_2 a^+ c_l ).
\end{array}
\end{equation}

\noindent The first four terms represent the free evolution of the
atom, the cavity field, the modes in port $1$ and $2$ respectively.
The last three terms represent the atom-cavity coupling, the
coupling of the cavity mode with the modes of port $1$ and with the
modes of port $2$. We can write the Heisenberg equations for each
operator

\begin{equation}
\begin{array}{l}

\dot{S}_-=-i\omega_0 S_- -2 \Omega S_z a \\

\dot{S}_z= \Omega (S_+ a + a^+ S_-) \\

\dot{a}=-i(\omega_0 + \delta) a - \Omega S_- + g_1\sum_k b_k + g_2\sum_l c_l\\

\dot{b_k}=-i\omega_k b_k + i g_1 a\\

\dot{c_l}=-i\omega_l c_l + i g_2 a   .\\

\end{array}
\end{equation}

\noindent We find for $t>t_0$, where $t_0$ is a reference of time

\begin{equation}\label{laplace}
\begin{array}{l}
b_k(t)=b_k(t_0)e^{-i\omega_k(t-t_0)}+i g_1 \int_{t_0}^t du~a(u)
e^{-i\omega_k(t-u)}\\
c_l(t)=c_l(t_0)e^{-i\omega_l(t-t_0)}+ i g_2 \int_{t_0}^t du~a(u)
e^{-i\omega_k(t-u)}  .
\end{array}
\end{equation}

\noindent Equations~(\ref{laplace}) are then injected in the
evolution equation for the cavity mode. For each mode $b_k$ and
$c_l$, the last term describes the field radiated by the cavity (\og
sources field\fg) and is responsible for the cavity damping. The
first term describes the free evolution and is responsible for the
noise in the quantum Langevin equation. Following Gardiner and
Collett~\cite{Gardiner85}, we define the input field in each port

\begin{equation}
\begin{array}{l}
b_{in}(t) = {\displaystyle \frac{1}{\sqrt{\tau}}\sum_k b_k(t_0)e^{-i\omega_k(t-t_0)}}\\
b_{in}^{'}(t) = {\displaystyle \frac{1}{\sqrt{\tau}}\sum_l
c_l(t_0)e^{-i\omega_l(t-t_0)}},
\end{array}
\end{equation}

\noindent where $\tau$ is defined by
\begin{equation}
\sum_k e^{-i\omega_k t}=\delta (t)\tau .
\end{equation}

\noindent The quantity $\tau$ has the dimension of a time and
depends on the mode density, which is supposed to be the same in
each port. The quantity $b^+_{in}b_{in}(t)$ (resp
${b^{'}}^+_{in}b^{'}_{in}(t)$) scales like a photon number per unit
of time and represents the incoming power in port $1$ (resp $2$).
Summing equations~(\ref{laplace}) over all modes in each port we
have

\begin{equation}\label{inout}
\begin{array}{l}
\sum_k b_k(t)=\sqrt{\tau} b_{in}(t) + i \frac{g_1}{2}\tau a(t)\\
\sum_l c_l(t)=\sqrt{\tau} b^{'}_{in}(t) + i \frac{g_2}{2}\tau a(t) .
\end{array}
\end{equation}

\noindent In the same way we define the reflected and transmitted
fields, for $t<t_0$

\begin{equation}
\begin{array}{l}
b_{r}(t) = {\displaystyle \frac{1}{\sqrt{\tau}}\sum_k b_k(t_0)e^{-i\omega_k(t-t_0)}}\\
b_{t}(t) = {\displaystyle \frac{1}{\sqrt{\tau}}\sum_l
c_l(t_0)e^{-i\omega_l(t-t_0)}},
\end{array}
\end{equation}

\noindent and in the same way we obtain
\begin{equation}\label{inout2}
\begin{array}{l}
\sum_k b_k(t)=\sqrt{\tau} b_{r}(t) -i\frac{g_1}{2}\tau a(t)\\
\sum_l c_l(t)=\sqrt{\tau} b_{t}(t) -i\frac{g_2}{2}\tau a(t) .
\end{array}
\end{equation}

\noindent  We suppose for simplicity that the coupling to each port
has the same intensity, $g_1=g_2$ which corresponds to the case of a
symmetric Fabry-Perot cavity. From equations~(\ref{inout}) and
(\ref{inout2}) we can easily derive the input-output equations for
the two-ports cavity

\begin{equation}
\begin{array}{l}
b_{r}(t) = b_{in}(t)+i \sqrt{\kappa}a\\
b_{t}(t) = b_{in}^{'}(t)+i \sqrt{\kappa}a  .  \\
\end{array}
\end{equation}

\noindent where we have taken $\kappa = |g_1|^2 \tau$. The evolution
equation for $a$ becomes

\begin{equation}
\dot{a}=-i(\omega_0+\delta) a  -\kappa a -\Omega S_- +
 i \sqrt{\kappa} b_{in} + i \sqrt{\kappa}
b_{in}^{'}.
\end{equation}

\begin{figure}[h,t]
\begin{center}
\includegraphics[height=5cm]{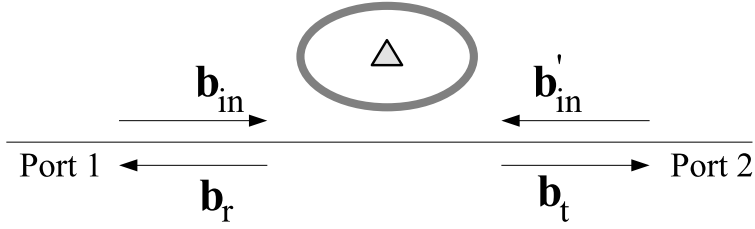}
\caption{\it Scheme of a cavity coupled quantum-dot system where the
cavity is evanescently coupled to ports $1$ and $2$. The incoming
field in port $1$ in now entirely transmitted if the coupling with
the cavity is switched off. This situation can describe a microdisk
cavity evanescently coupled to a waveguide. } \label{fig:evanescent}
\end{center}
\end{figure}

\noindent Note that this choice of definitions for the reflected and
transmitted field depends on the geometry of the problem. In the
situation depicted on figure~\ref{fig:transmission}, the incoming
field in port $1$ is entirely reflected if the coupling with the
cavity is switched off. In the case of a cavity evanescently coupled
to ports $1$ and $2$ (see figure~\ref{fig:evanescent}) the incoming
field in port $1$ would be entirely transmitted if the coupling with
the cavity were switched off. The definitions of $b_r$ and $b_t$
should just be inverted to describe this new situation. The theory
can also easily be adapted to the case of multiport cavities like
add-drop filters~\cite{akahane05}. The Heisenberg equations for the
cavity mode and the atomic operators are finally written in the
frame rotating at the drive frequency $\omega$

\begin{equation}\label{atome-in-cav}
\begin{array}{l}

\dot{S}_-=-i\Delta\omega S_- -2 \Omega S_z a \\

\dot{S}_z= \Omega (S_+ a + a^+ S_-) \\

\dot{a}=-i(\Delta\omega+\delta) a  -\kappa a -\Omega S_- + i\sqrt{\kappa} b_{in} + i\sqrt{\kappa} b_{in}^{'}\\

b_{r} = b_{in}+i\sqrt{\kappa} a \\

b_{t} = b_{in}^{'}+i\sqrt{\kappa} a  .

\end{array}
\end{equation}

\noindent Here $\Delta \omega = \omega_0 - \omega$. These equations
are the quantum coupled-mode equations for the evolution of the atom
and the cavity, driven by the external fields $b_{in}$ and
$b_{in}^{'}$. At this stage we shall suppose that the cavity
exchanges energy much faster with the input/output ports than with
the atom, that is : $\kappa \gg \Omega$. This regime is often called
the \emph{bad cavity regime} and we will from now on restrict
ourselves to that case. Note that the opposite case ($\Omega \gg
\kappa$) corresponds to the \emph{strong coupling regime} in which
the emission of a photon by the atom is coherent and reversible,
giving rise to the well-known phenomenon of quantum Rabi
oscillation~\cite{brunerabi}.

In the bad cavity regime, for a fixed frequency of the driving
field, the cavity mode can be adiabatically eliminated from the
equations, which means that we can take $\dot{a}=0$ at each time of
the system evolution. This implies for operator $a$

\begin{equation}
a=\frac{-\Omega S_-+i\sqrt{\kappa}(b_{in}+b_{in}^{'})}{i(\Delta
\omega+\delta)+\kappa} .
\end{equation}

The set of equations~(\ref{atome-in-cav}) becomes then

\begin{equation}\label{q-bloch}
\begin{array}{l}
\dot{S}_-={\displaystyle -i\Delta \omega S_-
-\frac{\Gamma}{2}t_0(\Delta\omega)S_- +
i\sqrt{\frac{\Gamma}{2}}(-2S_z)(b_{in}+b_{in}^{'})t_0(\Delta\omega)}\\

\dot{S}_z={\displaystyle -\Gamma \Re
\left(t_0(\Delta\omega)\right)\left(S_z+\frac{1}{2}\right)+\sqrt{\frac{\Gamma}{2}}\left(
i S_+(b_{in}+b_{in}^{'})t_0(\Delta\omega)+hc\right)}\\

b_t = {\displaystyle b_{in}^{'} \left( 1-t_0(\Delta\omega) \right)-
b_{in}t_0(\Delta\omega)
-i\sqrt{ \frac{\Gamma}{2} } S_- t_0(\Delta\omega)}\\

b_r = {\displaystyle b_{in} \left( 1-t_0(\Delta\omega) \right) -
b_{in}^{'} t_0(\Delta\omega) -i \sqrt{ \frac{\Gamma}{2} } S_-
t_0(\Delta\omega)} .
\end{array}
\end{equation}

\noindent We have introduced the relaxation time of the dipole in
the cavity mode $\Gamma=2\Omega^2/\kappa$. We have denoted
$t_0(\Delta\omega)$ the quantity ${\displaystyle 1/(1+i(\Delta\omega
+ \delta)/\kappa)}$. It will be shown in the next section that
$-t_0(\Delta \omega)$ corresponds to the transmission of an empty
cavity. Equations~(\ref{q-bloch}) hold between \emph{operators} :
they are quantum equivalents for the well-known optical Bloch
equations. They describe the effective interaction of a two-level
system with a one-dimensional continuum, mediated by a cavity : this
situation is generally referred to as the "one-dimensional
atom"~\cite{turchette95B}. In section III, we study this optical
medium in two regimes : the linear regime where the incoming field
is not strong enough to saturate the two-level system, and the
non-linear regime which we will study within the semi-classical
frame.

\section{Optical features of the one-dimensional atom}

In this part of the paper we focus on the optical behavior of the
one-dimensional atom. In particular, we define and compute a
transmission function for this medium, which shows two striking
features : first, in the linear regime, the presence of the dipole
induces a thin dip in the transmission function, leading to the
total reflection of the incident light (\emph{dipole induced
reflection}). Second, if the intensity of the driving field
increases, the transmission function shows a non-linear jump, the
switch happening for very low intensities of the driving field
(giant non-linear medium).

\subsection{Linear regime : dipole induced reflection}

In this part of the work, we suppose that the incoming field is very
weak, so that the saturation of the two-level system can be
neglected~:~the atomic population remains in the state $\ket{g}$,
and we can replace $S_z$ by its mean value $ \langle S_z \rangle
\approx -1/2$. Another way of introducing this approximation
consists in noting that the behavior of a two-level system in a
field containing very few excitations (zero or one photon) cannot be
distinghished from the behavior of the two lower levels of a
harmonic oscillator. $S_+$ and $S_-$, which are analogous to
creation and annihilation operators, should then have bosonic
commutation relation. Given that $[S_-,S_+]=-2S_z$, this condition
is fulfilled if $S_z\approx -1/2$. It is shown in appendix A that
$b_r$ and $b_t$ are related to $b_{in}$ and $b_{in}^{'}$ up to a
global phase by a unitary transformation, the scattering matrix
${\cal S}$ checking

\begin{equation}
\label{diffu} \left(
\begin{array}{c}
b_r \\
b_t
\end{array}
\right)= {\cal S} \left(
\begin{array}{c}
b_{in} \\
b_{in}^{'}
\end{array}
\right)= \frac{1}{1+i\zeta}\left(
\begin{array}{c c}
i\zeta & -1 \\
-1 & i\zeta
\end{array}
\right) \left(
\begin{array}{c}
b_{in} \\
b_{in}^{'}
\end{array}
\right),
\end{equation}

\noindent with

\begin{equation}
\zeta=\frac{\Delta \omega +
\delta}{\kappa}-\frac{\Gamma}{2\Delta\omega} .
\end{equation}

\noindent The system acts like a beamsplitter whose coefficients
depend on the frequency of the incoming fields. The statistics is
preserved by this transformation. If there is one photon of
frequency $\omega$ in the input field, the output field will be a
coherent superposition of a transmitted and a reflected photon of
frequency $\omega$, the amplitude of each part of the superposition
corresponding to the coefficients of the diffusion
matrix~(\ref{diffu}) as studied by Fan~\cite{fan05}. If the incoming
field is quasi-classical, the outcoming field will be
quasi-classical too and the reflection and transmission coefficients
can be interpreted in the usual way. We consider the transmission
coefficient in amplitude $t(\Delta \omega)={\cal S}_{12}={\cal
S}_{21}$ which reads

\begin{equation}
t(\Delta \omega)={\displaystyle \frac{-1}{1+i\zeta}}.
\end{equation}

\noindent As mentionned previously, the transmission of the empty
cavity, corresponding to $\Gamma=0$, fulfills

\begin{equation}\label{tcav}
t(\Delta\omega)=\frac{-1}{1+i{\displaystyle \frac{\Delta \omega +
\delta}{\kappa}}} = -t_0(\Delta \omega)\;.
\end{equation}

\begin{figure}[h,t]
\begin{center}
\includegraphics[height=7cm]{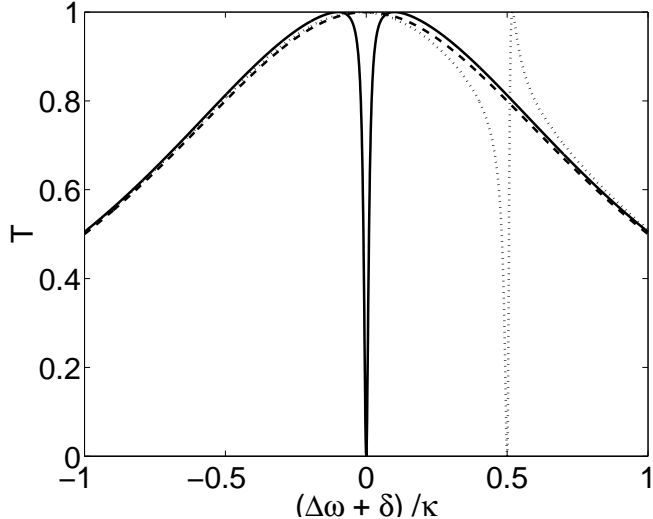}
\caption{\it Transmission of the optical system as a function of the
normalized detuning $(\Delta \omega + \delta)/ \kappa$ between the
cavity and the driving frequency. The curves are plotted with
$\Gamma/\kappa=1/500$. Dashed~:~transmission of the empty cavity.
Solid : transmission of the coupled quantum dot-cavity system, total
reflection is induced by the dipole. Dots : transmission of the
coupled quantum dot-cavity system with $\delta=-0.5\kappa$, the
signal is typical for a Fano resonance.} \label{fig:trans}
\end{center}
\end{figure}

\noindent The transmission coefficients in energy $T(\Delta
\omega)=|t(\Delta \omega)|^2$ and $T_0(\Delta \omega)=|t_0(\Delta
\omega)|^2$ are represented on figure~\ref{fig:trans} as functions
of the normalized detuning between the cavity and the driving field
$(\Delta \omega + \delta)/\kappa$. We took $\Gamma=\kappa/500$ which
fills the bad cavity regime condition. If there is no atom in the
cavity, $T_0(0)=1$ and the field is entirely transmitted at
resonance. If there is one resonant atom in the cavity, $T(0)=0$ and
the field is totally reflected by the optical system which behaves
as a frequency selective perfect mirror as evidenced by
Fan~\cite{fan05}. This \emph{dipole induced reflection,} reminiscent
of dipole induced transparency evidenced by Waks et
al.~\cite{waksprl}, cannot be attributed to a phase-shift induced by
the atom, putting the cavity out of resonance. On the contrary, it
is due to a totally destructive interference between the incoming
field and the field radiated by the dipole as it appears on
equation~(\ref{interf}):

\begin{equation}\label{interf}
b_t=-\left[b_{in}+ i\sqrt{\frac{\Gamma}{2}}S_-\right],
\end{equation}

\noindent the stationary state of the atomic dipole being

\begin{equation}
S_-=i\sqrt{\frac{2}{\Gamma}}b_{in} .
\end{equation}

\noindent The global $-$ sign in equation~(\ref{interf}) is due to
the cavity resonance. The interference is destructive because the
fluorescence field emitted by a two-level system is phase-shifted
by $\pi$ with respect to the driving field as pointed out by
Kojima~\cite{kojima04}. If the dipole is not resonant with the
cavity the transmission is a Fano resonance as underlined by
Fan~\cite{fan05}.

\noindent If $\delta=0$, $T$ reads

\begin{equation}
T(\Delta \omega)=\frac{1}{1+\left( {\displaystyle \frac {\Gamma}{2
\Delta \omega } -\frac{\Delta \omega}{\kappa}}  \right)^2}\;.
\end{equation}

\noindent The dip linewidths can be easily computed from the
solutions of the equation $T=1/2$. Remembering that $\Gamma \ll
\kappa$, we find that the linewidth of the broadest transmission
peak is the cavity linewidth

\begin{equation}
\Delta \omega_{1/2} = \kappa  \; ,
\end{equation}

\noindent whereas the linewidth of the narrow dip is corresponds to
the linewidth of the atom dressed by the cavity mode

\begin{equation}
\delta \omega_{1/2} = \Gamma .
\end{equation}

\noindent It appears that in the linear regime, the one-dimensional
atom is a highly dispersive medium which can be used to slow down
light as it is realized using media showing Electromagnetically
Induced Transparency. This effect is studied in appendix C using a
model including leaks.

\subsection{Non-linear regime : giant optical non-linearity}

We are now interested in the optical behavior of the one-dimensional
atom for arbitrary intensities of the incoming field. Following
Allen and Eberly~\cite{Allen}, we adopt the semi-classical
hypothesis where the quantum correlations between atomic operators
and field operators can be neglected. We shall comment the range of
validity of this approximation at the end of this section. We take
the mean value of equations~(\ref{q-bloch}) to obtain relations
between the quantities $\langle b_{in} \rangle$, $\langle b_t
\rangle$, $ \langle b_r \rangle$ as they could be measured using a
homodyne detection. In the following of this paper we shall take $
\langle b_{in}^{'}\rangle=0$. Writing $s= \langle S_- \rangle$,
$s_z= \langle S_z \rangle$, and identifying $b_t$ (respectively
$b_r$ and $b_{in}$) to $\langle b_t \rangle$ (respectively to $
\langle b_r \rangle$ and $\langle b_{in} \rangle$) we obtain

\begin{equation}\label{bloch-optique}
\begin{array}{l}
\dot{s}={\displaystyle -i\Delta \omega s -\frac{\Gamma}{2
}t_0(\Delta\omega)s + i \sqrt{\frac{\Gamma}{2}}
(-2 s_z) b_{in}t_0(\Delta\omega)}\\

\dot{s}_z={\displaystyle -\Gamma
\Re\left(t_0(\Delta\omega)\right)\left(s_z+\frac{1}{2}\right)+
\sqrt{\frac{\Gamma}{2}}\left(i s^*b_{in}t_0(\Delta\omega)+cc\right)}\\

b_t = -{\displaystyle \left( b_{in} + i \sqrt{\frac {\Gamma}{2}}s \right) t_0(\Delta\omega)}\\

b_r = b_{in}+b_t \;.

\end{array}
\end{equation}

\noindent Equations~(\ref{bloch-optique}) are similar to the well
known Bloch optical equations for a two-level system interacting
with a classical field with a coupling constant $\Gamma$.
Nevertheless, in this case the dipole relaxation rate is related
to the coupling constant, whereas usually the two parameters are
independant. This is due to the fact that the dipole is driven and
relaxes via the same ports $1$ and $2$. We obtain after some
little algebra detailed in appendix B the stationary solution for
the population of the two-level system

\begin{equation}
\begin{array}{l}
s={\displaystyle
\sqrt{\frac{2}{\Gamma}}\frac{1}{1+x}\frac{ib_{in}}{1+{\displaystyle
\frac{2i \Delta \omega}{\Gamma t_0(\Delta\omega)}}} }\\

s_z={\displaystyle -\frac{1}{2} \frac{1}{1+x}}  \; , \\
\end{array}
\end{equation}

\noindent where we have introduced the saturation parameter $x$

\begin{equation}
x={\displaystyle \frac{|b_{in}|^2}{P_c(\Delta \omega)}}.
\end{equation}

\noindent $P_c(\Delta \omega)$ is the critical power necessary to
reach $s_z=-1/4$, satisfying

\begin{equation}
\begin{array}{l}
P_c(\Delta \omega)={\displaystyle \frac{\Gamma}{4}\phi(\omega)}\\

\phi(\omega)={\displaystyle \left(\frac{2\Delta \omega}{\Gamma}\right)^2+
\left(\frac{2\Delta \omega}{\Gamma} \frac{\Delta \omega + \delta}{\kappa}-1\right)^2} \; . \\
\end{array}
\end{equation}

\noindent $P_c$ scales like a number of photons per second. At
resonance it corresponds to one forth of photon per lifetime. Out
of resonance it is increased by a factor $\phi(\omega)$ which can
be seen as the inverse of an adimensional cross-section. We define
an adimensional susceptibility $\alpha$ for the two-level system
\begin{equation}
s=\sqrt{\frac{2}{\Gamma}}\alpha b_{in} \;,
\end{equation}

\noindent  where $\alpha$ reads

\begin{equation}
\alpha = \frac{1}{1+x}\frac{i}{1+{\displaystyle \frac{2i \Delta
\omega}{\Gamma t_0(\Delta \omega)}} } \;.
\end{equation}

\begin{figure}[h,t]
\begin{center}
\includegraphics[height=8cm]{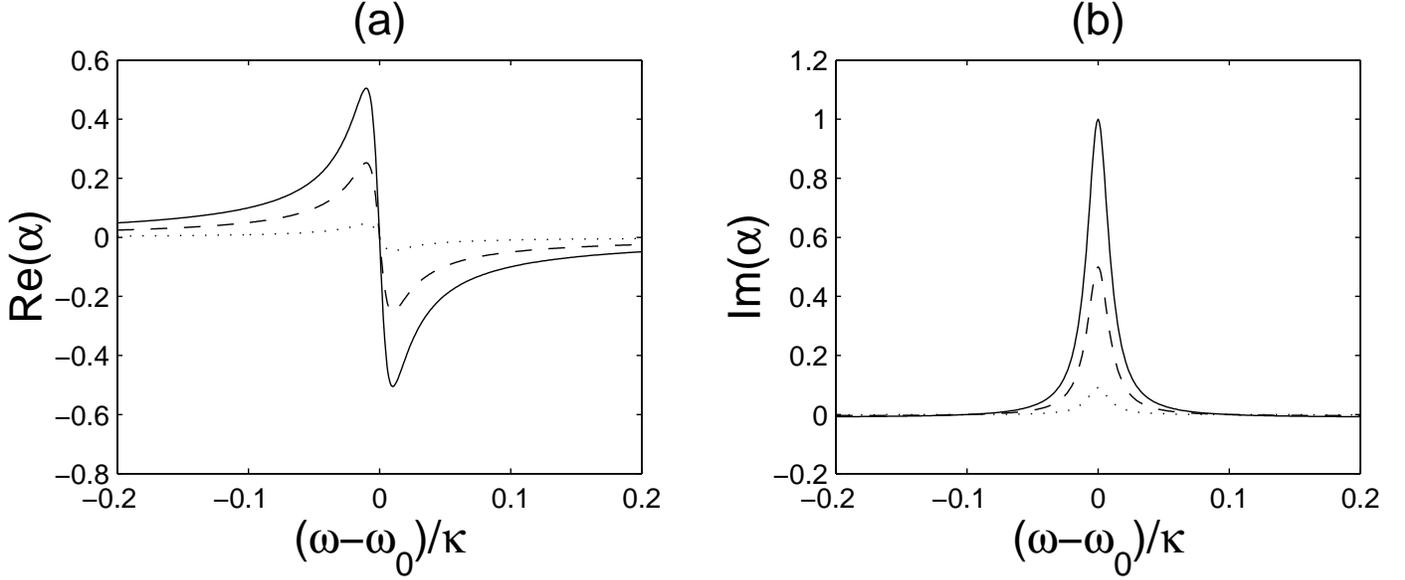}
\caption{\it Susceptibility $\alpha$ of the atomic dipole as a
function of $(\omega-\omega_0)/\kappa=-\Delta \omega /\kappa$, for
different values of the saturation parameter $x$. (a) Real part of
$\alpha$. (b) Imaginary part of $\alpha$. Solid : x=0. Dashed : x=1.
Dots : x=10.} \label{fig:alpha}
\end{center}
\end{figure}

\noindent We have plotted in figure~\ref{fig:alpha} the evolution of
the real and imaginary part of the susceptibility as a function of
$\omega-\omega_0=-\Delta \omega$ for different values of the
saturation parameter. As expected, the sensitivity to the incoming
field's intensity, that is the non-linear effect, is maximal for
$\Delta \omega=0$ and $\alpha$ checks

\begin{equation}
\alpha=\frac{i}{1+x} \;.
\end{equation}

\begin{figure}[h,t]
\begin{center}
\includegraphics[height=6cm]{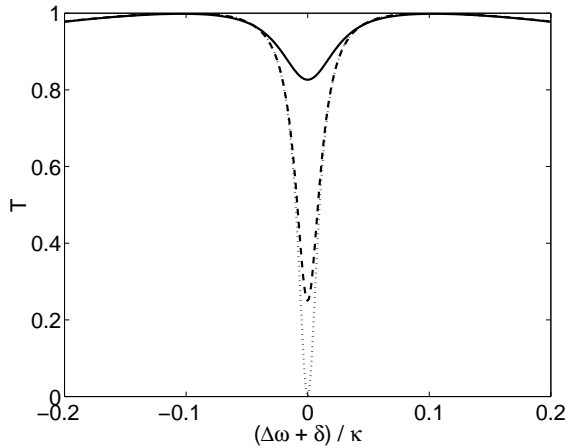}
\caption{\it Transmission of the optical system as a function of the
normalized detuning $(\Delta \omega + \delta)/ \kappa$ between the
quantum dot and the driving frequency for different values of
saturation parameter at resonance $x=4|b_{in}|^2/\Gamma$. We took
$\delta=0$ for convenience. Dots : $x=0$. Dashed-dot : $x=1$. Solid
: $x=10$.}\label{fig:nl-ideal}
\end{center}
\end{figure}

\noindent At resonance $\alpha$ is purely imaginary : the field is
entirely absorbed by the dipole. The behavior of the two-level
system drastically changes from $|b_{in}|^2\sim 0$ to
$|b_{in}|^2\sim 10 P_c$ which corresponds to a very low switching
value. Any two-level system is then a giant optical non-linear
medium. In the specific case of the one-dimensional atom, the
fluorescence field interferes with the driving field, and a
signature of the giant non-linearity can be observed in the output
field. We have represented in figure~\ref{fig:nl-ideal} the
transmission coefficient $T=|t(\Delta \omega)|^2$ for different
values of the incoming power. For low values the system is not
saturated and the dipole blocks the light. For $|b_{in}|^2=P_{in}>10
P_c$ the dipole is saturated and cannot prevent light from crossing
the cavity. This non-linear behavior is obvious if we restrict
ourselves to the resonant case. At resonance indeed the transmission
and reflection coefficients in amplitude $t$ and $r$ write

\begin{equation}
\begin{array}{l}
t={\displaystyle \frac{-x}{1+x}}\\
r={\displaystyle \frac{1}{1+x}}  \; ,
\end{array}
\end{equation}

\noindent which implies for the transmitted and reflected power
$P_t$ and $P_r$

\begin{equation}
\begin{array}{l}
P_t={\displaystyle \frac{x^2}{(1+x)^2} P_{in}}\\
P_r={\displaystyle \frac{1}{(1+x)^2} P_{in}}  \;.
\end{array}
\end{equation}

\begin{figure}[h,t]
\begin{center}
\includegraphics[height=10cm]{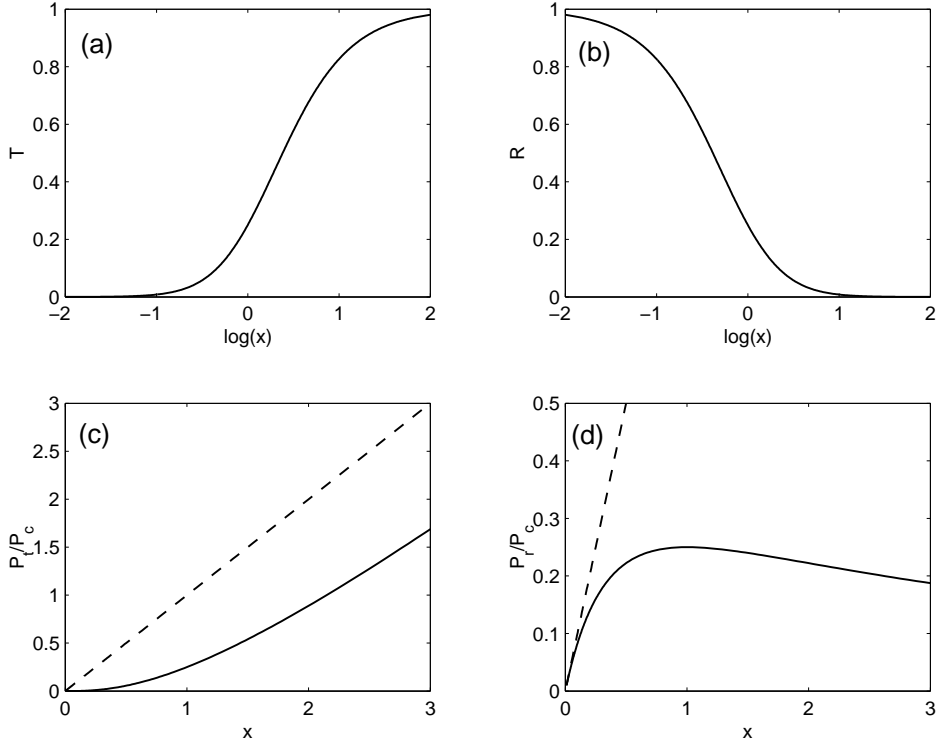}
\caption{\it $(a)$ Transmission, $(b)$ reflection coefficient as a
function of the logarithm of the saturation parameter on resonance
$\log(x)=\log(4|b_{in}|^2/\Gamma)$. $(c)$ Solid : normalized
transmitted power $P_t/P_c$ and $(d)$ normalized reflected power
$P_r/P_{c}$ as a function of the saturation parameter $x$. Dashed :
normalized incoming power $P_{in}/P_c$. The transmitted field
corresponds to the driving field lowered by one photon per lifetime,
which has been absorbed by the atom. This reflected field increases
with the driving field until $x=1$, and then decreases because of
the saturation of the two-level system.} \label{fig:nl-ideal-res}
\end{center}
\end{figure}

\noindent $R$, $T$, $P_r$ and $P_t$ are plotted in
figure~\ref{fig:nl-ideal-res}. As expected a non-linear jump in
the transmission coefficient happens at a typical power for the
incoming field $P_{in}\sim P_c/2$. Note that this giant optical
non-linearity has been pointed out in the case of a two-level
system in an asymmetric cavity~\cite{hofmann03}, the non-linear
jump being observable in the phase of the reflected field.

\noindent It appears that $P_r+P_t\neq P_{in}$ even for an ideal
non-leaky system as considered in this section. To understand this,
let us remind that $P_t+P_r$ is the power of the coherently diffused
field, which is predominent if the driving field is weak. On the
contrary, when the dipole is saturated, the fluorescence field is
emitted with a random phase and cannot interfere with the driving
field anymore~\cite{cohen,hofmann03}. This incoherent diffusion
process is responsible for a noise whose power $P_{noise}$ allows to
preserve energy conservation

\begin{equation}
P_{noise}=P_{in}-P_r-P_t \sim \frac{2x}{(1+x)^2}P_{in}  \;.
\end{equation}

\begin{figure}[h,t]
\begin{center}
\includegraphics[height=7cm]{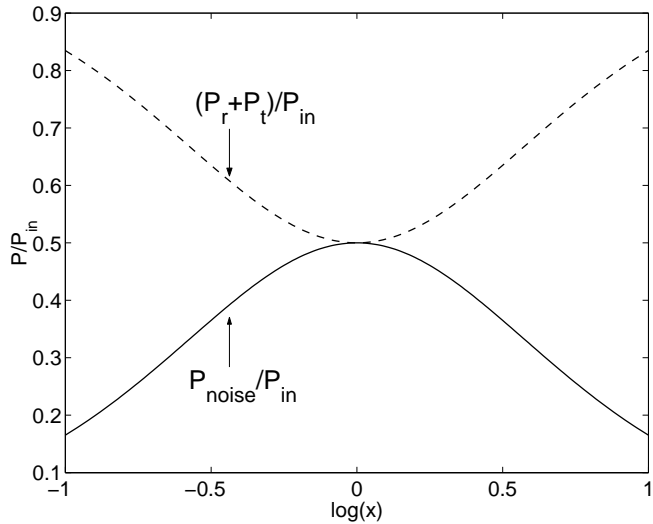}
\caption{\it $\log(P/P_{in}$) as a function of $\log(x)$. Solid :
$\log(P_{noise}/P_{in})$. Dashed : $\log(P_{r}+P_t/P_{in})$.
}\label{fig:noise}
\end{center}
\end{figure}

\noindent Let us mention that $P_{noise}$ could be detected with
direct photon counting and would be split between the two output
ports. We have plotted in figure~\ref{fig:noise} the relative
contribution of the noise power $P_{noise}$ and of the coherently
diffused fields $P_r+P_t$ over the incoming power $P_{in}$, as a
function of the logarithm of the saturation parameter. The noise
contribution is maximal for $x=1$. This also gives us a glimpse of
the range of validity for the semi-classical assumption, which
correctly describes the problem only out of the non-linear jump.

\subsection{Quantifying the giant non-linearity}

As underlined before, the non-linearity is giant because of two main
effects, which are caracteristics of the one-dimensional atom
geometry: first, any photon that is sent in the input field reaches
the single two-level system; second, the fluorescence field is
entirely directed in the output ports, so that there are no leaks
and we can operate at resonance. To quantify the non-linearity it is
convenient to observe that the transmission and reflection jumps
could be obtained using an optical medium inducing a non-linear
phase jump of $\pi$ without absorption, the jump happening for a
typical intensity $I_\pi \sim 10 P_c/\sigma$ where $\sigma$ is the
surface on which light is focused and the factor of $10$ is
evaluated from figure \ref{fig:nlresleak}. Let us compute the
typical intensity in our case. The critical power $P_c$ is one forth
photon per lifetime, that is, with a wavelength $\lambda\sim 1\mu m$
and a lifetime $\tau\sim 100$~ps which correspond to realistic
experimental parameters as it will appear in section V, $P_c\sim
1$~nW. We shall take $\sigma\sim 10^{-8}$~cm$^2$ which corresponds
to the typical surface of a semiconducting microcavity. We obtain
$I_\pi\sim 1$W/cm$^2$. Let us consider a non-linear Kerr medium with
a refractive index given by $n=n_0+n_2I$ where $I$ is the intensity
of the light beam crossing the medium. The non-linear phase-shift
acquired by the beam is

\begin{equation}
\phi_{nl}=\frac{2\pi}{\lambda}Ln_2 I .
\end{equation}

\noindent Given that the non-linear index of bulk semiconductor
(like GaAs) at half gap excitation is typically
$n_2=10^{-13}$~cm$^2$/W~\cite{said}, the length of medium should
be $5. 10^3$ km to reach a $\pi$ phase shift with the same
intensity ! Resonant experiment using an atomic vapor in low
finesse cavity have reached values of $n_2\sim 10^{-7}$~cm$^2$/W
while preserving a quantum noise limited operation~\cite{Gra98} :
a $\pi$ phase shift could be obtained after $5$~m of vapor. More
recently there has been work on slow light using
electromagnetically induced transparency exhibiting giant resonant
non linear refractive index $n_2=0.18$~cm$^2$/W ~\cite{Hau99},
leading to a length of a few mm to reach the same effect.

\section{Influence of the leaks}

In section III we have seen that a one-dimensional atom driven by a
low intensity field is a highly dispersive medium that could be used
to slow down light as it is shown in appendix C. Morover, if this
medium is driven by a resonant field, its transmission shows a
non-linear jump at a very low switching intensity. We aim at
observing these two effects using solid state two-level systems and
cavities. In order to prepare the feasibility study which will be
held in the next section, we focus in this part of the paper on the
quantitative influence of the leaks on the transmission function of
the system. We note $\gamma_{at}$ and $\gamma_{cav}$ the leaks from
the atom and from the cavity respectively. Given that we will deal
with artificial atoms such as quantum dots, we shall also consider
the excitonic dephasing $\gamma^*$. The set of
equations~(\ref{atome-in-cav}) becomes

\begin{equation}\label{leaks}
\begin{array}{l}
\dot{S}_-=-i\Delta\omega S_- - 2 \Omega S_z  a - {\displaystyle \frac{\gamma_{at}}{2}}S_- -\gamma^* S_- + G\\

\dot{S}_z= \Omega (S_+ a + a^+ S_-) - \gamma_{at}(S_z+1/2) + K \\

\dot{a}=-i(\Delta\omega+\delta) a  -\kappa a -\Omega S_- +i\sqrt{\kappa} b_{in}+i\sqrt{\kappa} b_{in}^{'} - {\displaystyle \frac{\gamma_{cav}}{2}} a  + H\\

b_t = b_{in}^{'}+i\sqrt{\kappa} a \\

b_r= b_{in}+i\sqrt{\kappa} a \;.

\end{array}
\end{equation}

\noindent $K$, $G$ and $H$ are noise operators due to the
interaction of the atom and the cavity with their respective
reservoirs, respecting $<G>=<H>=<K>=0$. The noise prevents us from
obtaining relations between incoming and outcoming field operators.
As a consequence, even in the linear case, we will deal with
expectation values of the fields as they could be obtained in a
homodyne detection experiment.

\subsection{Linear regime}

First we consider the linear case, so that $<S_z> \approx -1/2$.
Using the same notations as in the previous section, we obtain after
adiabatic elimination of the cavity mode

\begin{equation}
\begin{array}{l}\label{equ}
\dot{s}=- i\Delta \omega s -{\displaystyle
\frac{\Gamma}{2}\frac{Q}{Q_0}\left[t^{'}_{0}+\frac{Q_0}{Q}\frac{\gamma_{at}+2\gamma^*}{\Gamma}\right]
s +
i \frac{Q}{Q_0}\sqrt{\frac{\Gamma}{2}}b_{in}t^{'}_0}\\
b_t={\displaystyle - \frac{Q}{Q_0} t^{'}_{0} b_{in} - i
\frac{Q}{Q_0} \sqrt{\frac{\Gamma}{2}}t^{'}_{0}s }
\\
b_r=b_{in}+b_t
 \;.
\end{array}
\end{equation}

\noindent We have introduced the adimensional quantity $t_0^{'}$
such as

\begin{equation}
t^{'}_{0}(\Delta \omega)={\displaystyle \frac{1}{1+{\displaystyle i\frac{Q}{Q_0}\frac{\Delta\omega+\delta}{\kappa}}}}\\
 \; .
\end{equation}

\noindent The parameter $Q_0$ is  the quality factor of the cavity
mode due to the coupling with the one-dimensional continua of
modes. The parameter $Q$ is the total quality factor and includes
the coupling to leaky ones. $Q_0$ and $Q$ fulfill

\begin{equation}
Q_0/Q=1+\gamma_{cav}/2\kappa \;.
\end{equation}

\noindent If the dipole is non-leaky, that is if $\gamma_{at}=0$,
its relaxation rate in the cavity mode is equal to ${\displaystyle
\Gamma Q/Q_0}$. It is lower than in the case of a cavity perfectly
matched to the input and output modes, because the cavity being
enlarged, the density of modes on resonance with the dipole is
lower. It is convenient to define the ratio $f$

\begin{equation}
f= \frac{Q}{Q_0}\frac{\Gamma} {\gamma_{at}+2\gamma^*} \; .
\end{equation}

\noindent Note that the ratio $f$ is different from the Purcell
factor $F_p$~\cite{purcell46} of the two-level system, defined
indeed as the spontaneous emission rate in the cavity mode over the
emission rate in the vacuum space, which we shall denote
$\gamma_{free}$. The quantities $f$ and $F_p$ are related by the
following equation

\begin{equation}\label{f}
f=\frac{\gamma_{free}}{\gamma_{at}+2\gamma^*}F_p.
\end{equation}

\noindent In the very simple case where $\gamma^*=0$ and
$\gamma_{at}=\gamma_{free}$, we have $f=F_p$. Note that the
excitonic dephasing $\gamma^*$ reduces the ratio $f$ and may lead to
the reduction of the contrast of the experimental signal. The
transmission coefficient of the empty cavity can be written $-Q/Q_0
t^{'}_0(\Delta\omega)$, the reflection coefficient being $r=1+t$. If
the cavity contains one atom, the transmission coefficient of the
system has the following expression

\begin{equation}
t(\Delta \omega)={\displaystyle \frac{Q}{Q_0} t^{'}_0 \left[-1+
\frac {f} {f+{\displaystyle
\left(\frac{i\Delta\omega}{\gamma_{at}+2\gamma^*}+1\right)\left(i\frac{Q}{Q_0}\frac{\Delta\omega+\delta}{\kappa}+1\right)}}
 \right]} \; ,
\end{equation}

\noindent It appears that the one-dimensional atom case requires
$Q/Q_0\sim1$, $(f,F_p)\rightarrow \infty$, which justifies for the
so-called "Purcell regime" we have referred to until now. At
resonance, the transmission and reflection coefficients in energy
for an empty cavity can be written

\begin{equation}\label{Tmax}
\begin{array}{l}
T_{max}={\displaystyle
\left(\frac{Q}{Q_0}\right)^2} \\
R_{min}={\displaystyle \left(1-\frac{Q}{Q_0}\right)^2},
\end{array}
\end{equation}

\noindent whereas if the cavity contains one resonant two-level
system, their expression become

\begin{equation}
\label{Tmin}
\begin{array}{l}
T_{min}={\displaystyle \left(\frac{Q}{Q_0}\right)^2
\left(\frac{1}{1+f}\right)^2} \\
R_{max}={\displaystyle \left(
1-\frac{Q}{Q_0}\frac{1}{1+f}\right)^2}.
\end{array}
\end{equation}

\begin{figure}[h,t]
\begin{center}
\includegraphics[height=10cm]{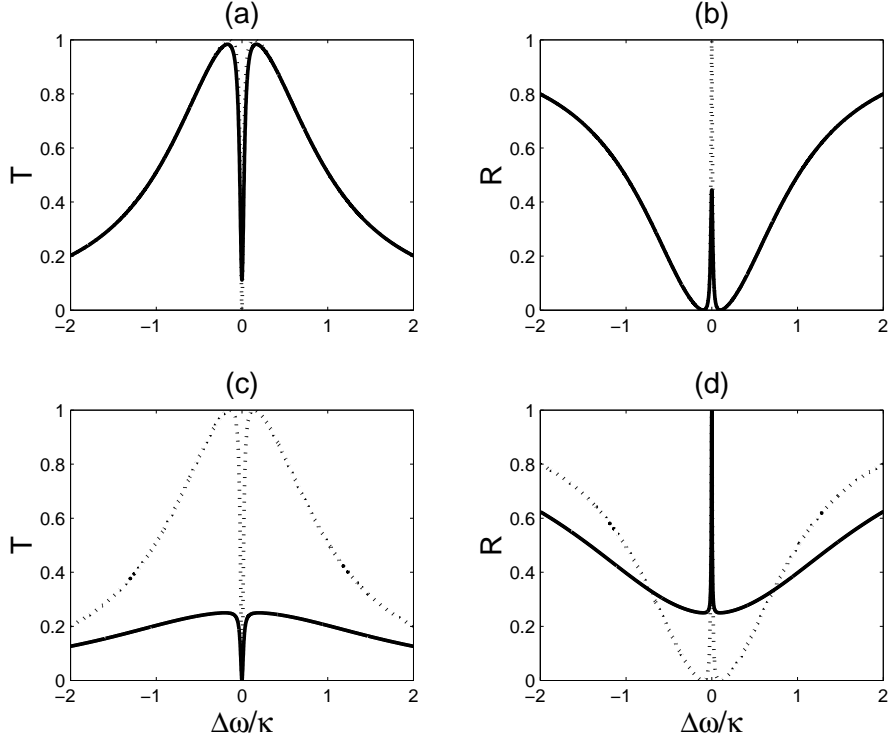}
\caption{\it Evolution of $T$ and $R$ as functions of the
atom-cavity detuning for different values of $Q$, $Q_0$ and $f$.
$\delta$ has been taken equal to $0$ for convenience. Dots : ideal
case with $Q=Q_0$ and $f\rightarrow \infty$. $(a)$ : $T$ with
$Q=Q_0$ and $f=2$. $(b)$ : $R$ with the same parameters. $(c)$ :
$T$ with $Q/Q_0=1/2$ and $f=\rightarrow \infty$. $(d)$ : $R$ with
the same parameters. } \label{fig:TR}
\end{center}
\end{figure}

\noindent We have plotted in figure~\ref{fig:TR} the evolution of
$T$ and $R$ as functions of the atom-cavity detuning for different
values of $Q$, $Q_0$ and $f$. The plots $(a)$ and $(b)$ correspond
to the case of a cavity perfectly connected to the input and output
mode ($Q=Q_0$) interacting with a leaky two-level system. On the
plots $(c)$ and $(d)$, we consider the case of an atom perfectly
connected to a leaky cavity mode ($f\rightarrow \infty$ and $Q/Q_0 <
1$). Note that the limit $f \rightarrow \infty$ can be taken without
reaching the strong coupling regime, provided the coupling to leaky
modes and the excitonic dephasing vanish
($\gamma_{at},\gamma^*\rightarrow 0$).

\noindent  Let us stress that the reflection can be total even if
the cavity is leaky. This apparently striking result is due to a
totally constructive interference between the driving field and the
field radiated by the optical system, which cannot be split into a
cavity and an atom, but must be considered as a whole. This feature
also appears on the normalized leaks on resonance ${\cal L}$ given
by $R+T=1-{\cal{L}}$, which fulfills

\begin{equation}
{\cal
L}=2\sqrt{R}\sqrt{T}=\frac{2Q}{Q_0}\frac{1}{1+f}\left(1-\frac{Q}{Q_0}\frac{1}{1+f}\right)
\end{equation}

\noindent The leaks can be approximated for $f \gg 1$ by the
following expression

\begin{equation}
{\cal L}\sim\frac{2Q}{Q_0f}=\frac{2\gamma_{at}}{\Gamma},
\end{equation}

\noindent which has a clear physical meaning : the leaks can be
interpreted as the rate of photons lost by the atom over the rate of
photons funneled in the output mode. This quantity decreases down to
$0$ when the atomic leaks become vanishingly small, even if the atom
is placed in a leaky cavity.

\subsection{Non-linear regime}

\noindent We consider now the case of a leaky optical system
described by equations~(\ref{leaks}). We shall restrict ourselves to
the resonant case and to the semi-classical hypothesis. For sake of
simplicity we shall also take $\gamma^*=0$, which is a realistic
hypothesis as it will be shown in the next section. As before we can
adiabatically eliminate the cavity from the equations. Using the
same definitions for $\Gamma$, $f$, $Q$ and $Q_0$, we establish the
optical Bloch equations for the leaky system

\begin{equation}
\begin{array}{l}
\dot{s}={\displaystyle
-\frac{\Gamma}{2}\frac{Q}{Q_0}\left(1+\frac{1}{f}\right)s+\sqrt{\frac{\Gamma}{2}}\frac{Q}{Q_0}(-2s_z)ib_{in}}\\
\dot{s}_z={\displaystyle -\Gamma\frac{Q}{Q_0}
\left(1+\frac{1}{f}\right)\left(s_z+\frac{1}{2}\right)+\sqrt{\frac{\Gamma}{2}}\frac{Q}{Q_0}(ib_{in}s^*+cc)}\\
b_t={\displaystyle
-b_{in}\frac{Q}{Q_0}-i\sqrt{\frac{\Gamma}{2}}\frac{Q}{Q_0}s}\\
b_r={\displaystyle
b_{in}\left(1-\frac{Q}{Q_0}\right)-i\sqrt{\frac{\Gamma}{2}}\frac{Q}{Q_0}s}
\;.

\end{array}
\end{equation}

\noindent At it is shown in appendix B, the stationary solutions can
be written

\begin{equation}
\begin{array}{l}
s_z={\displaystyle -\frac{1}{2}\frac{1}{1+x^{'}}}\\
s={\displaystyle
\sqrt{\frac{\Gamma}{2}}\frac{ib_{in}}{1+x^{'}}\frac{1}{1+{\displaystyle
\frac{1}{f}}}} \;.
\end{array}
\end{equation}

\noindent with modified values for the saturation parameter $x^{'}$
and the critical power $P_c^{'}$

\begin{equation}
\begin{array}{l}
x^{'}={\displaystyle |b_{in}|^2/P_{c}^{'}}\\
P_{c}^{'}={\displaystyle
\frac{\Gamma}{4\beta^2} }\;.  \\
\end{array}
\end{equation}

\noindent We have introduced the parameter $\beta={\displaystyle
\frac{f}{1+f}}$. The quantity $\beta^2$ can be seen as the
probability for a resonant photon sent in the input mode to be
absorbed by the optical system. The power necessary to saturate the
two-level system, that is to reach $s_z=-1/4$, is higher than in the
ideal case which is a natural consequence of the leaks. The
transmission coefficient in energy can be written

\begin{equation}
T={\displaystyle
\left(\frac{Q}{Q_0}\right)^2\left[\frac{\beta}{1+\beta^2
x}-1\right]^2} \;.
\end{equation}

\begin{figure}[h,t]
\begin{center}
\includegraphics[height=6cm]{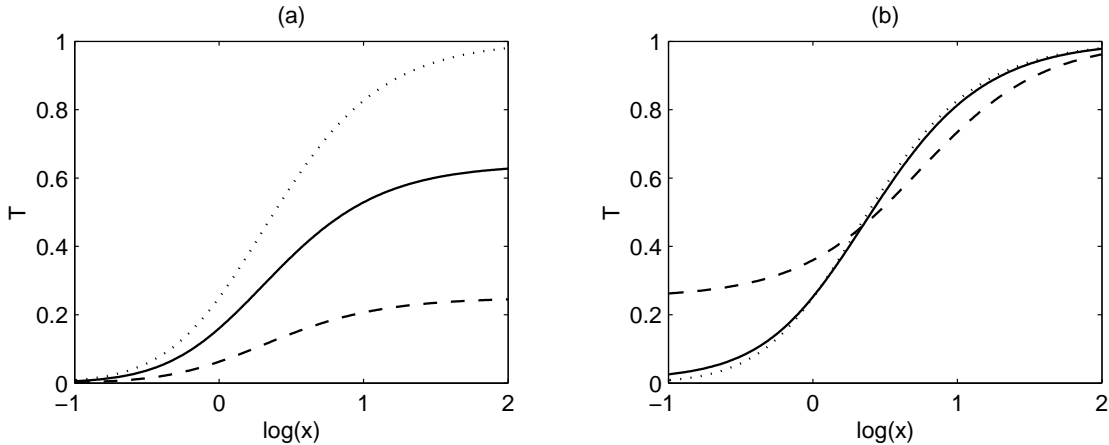}
\caption{\it Transmission of the optical system on resonance as a
function of the logarithm of the saturation parameter
$\log(x)=\log(4P_{in}/\Gamma)$. (a): We fixed $\beta=1$ which
corresponds to high values of $f$. Dots : $Q=Q_0=1000$ (ideal case).
Solid : $Q=800$. Dashed : $Q=500$. The obtained signals are the
ideal signal multiplied by $(Q/Q_0)^2$. (b): We took $Q=Q_0=1000$.
Dots : ideal case. Dashed : $f=1$. Solid : $f=10$. The non-linear
jump happens for higher values of the saturation parameter.
}\label{fig:satur2}
\end{center}
\end{figure}

\noindent We have plotted in figure~\ref{fig:satur2} the
transmission coefficient $T$ as a function of the saturation
parameter in the non-leaky case $x=4P_{in}/\Gamma$. The limit of the
signal for $x\rightarrow 0$ is $T_{min}$ because the two-level
system is not saturated. If $x\rightarrow \infty$ the signal tends
to $T_{max}$ : when the two-level system is saturated, the optical
system behaves like an empty cavity. On the left, we fixed $\beta=1$
which may be realized with high values of the ratio $f$, and we
considered different leaky cavities. In this case, the transmission
coefficient simply corresponds to the ideal transmission coefficient
multiplied by $(Q/Q_0)^2$. On the right, we have considered a
non-leaky cavity ($Q/Q_0=1$) and different values of the ratio $f$.
The jump happens for higher values of the saturation parameter,
which was expected.

\section{Feasibility study}

In the two previous sections, we have seen that a one-dimensional
atom, even leaky, induces the reflection of a low intensity driving
field, giving rise to a highly dispersive transmission pattern, and
behaves like a giant non-linear medium, with typical switching
intensities of one photon per lifetime. This section aims at showing
that these striking features can be observed using solid state
two-level systems and cavities. As a first step, we shall comment on
the validity of the two-level system model in the case of a single
exciton embedded in a quantum dot. As a second step, we will focus
on a well-known semi-conducting microcavity whose caracteristics
depend on a small set of easily adjustable parameters : the
micropillar. Micropillars are very good candidates for this
application because the light they emit is directional. As a
consequence they have already been used with success as single
photon sources~\cite{Solomon,Moreau01} and indistinguishable photon
sources~\cite{Var05}. We shall optimize these parameters in view of
observing the dipole induced reflection or the non-linear effect. As
a third step we will draw a comparison between the performances of
the device when it is operated as a single photon source or as a
giant non-linear medium.

\subsection{How good a two-level system is a semiconductor quantum dot ?}

Semiconductor quantum dots displaying a very high structural and
optical quality can be obtained using self-assembly in molecular
beam epitaxy~\cite{jmg2}. Such nanostructures confine both electrons
and holes on the few-nanometer scale, and support therefore a
discrete set of confined electronic states. In its ground state
$\ket{g}$, the quantum dot is empty, whereas the lowest bright
energy level $\ket{e}$ corresponds to the situation where it
contains one electron-hole pair called \emph{exciton}. Sharp atomic
like fluorescence~\cite{marzin} and absorption~\cite{karrai} lines,
associated to optical transitions between $\ket{e}$ and $\ket{g}$
can be observed experimentally.

As far as the spin structure is concerned, the projection of the
electronic spin on the growth axis of the dot is either $1/2$ or
$-1/2$ , whereas the projection of the hole's spin is either $3/2$
or $-3/2$ ("heavy holes"). This corresponds to four distinct spin
values for the exciton. However, only two excitons are coupled to
the ground state by the electromagnetic field, namely
$\ket{-1/2,3/2}$ and $\ket{1/2,-3/2}$ ("bright" excitons). The two
other excitons have a total spin projection of $2$ or $-2$. They
remain optically uncoupled or "dark" because of the selection rules
governing the dipolar electrical hamiltonian, and they don't have to
be taken into account.

For a quantum dot showing perfect cylindrical symmetry around its
growth axis, the two excitonic states are degenerate. In practice
the symmetry is not perfect and the exchange interaction splits the
doublet in two eigenstates, which are coupled to the ground state by
two orthogonally linearly polarized fields~\cite{bayer}. At this
point, a quantum dot appears as a "V-type" system rather than as a
two-level system. However, recent experiments have shown that in a
cryogenic environment and under resonant pumping, which will
correspond to our experimental conditions, electrons' and holes'
spins are frozen at the exciton lifetime scale~\cite{paillard}. As a
consequence, it is possible to work at a given linear polarisation
and to ignore the other excitonic spin state, allowing an effective
treatment of the quantum dot as a two-level system.

One could fear that pumping the quantum dot beyond its saturation
intensity may lead to the creation of two electron-hole pairs
(\emph{biexcitonic states}, denoted \emph{XX}). However, because of
the interaction between two excitons, the transition between
\emph{XX} and $\ket{e}$ is energetically different (a few meV
typically) from the transition between $\ket{e}$ and the fundamental
state $\ket{g}$. This effect allows to address spectrally the
excitonic state under interest : if the driving field is resonant
with the excitonic transition (which is the case in the
demonstration of the giant non-linearity) or slightly detuned from
the excitonic transition (which is the case in the experiments
aiming at showing the dipole induced reflection, where the detuning
is less than $1$~meV, corresponding to the spectral width of a
micropillar), \emph{XX} does not have to be taken into account.

The interaction of the exciton with the phonons of the surrounding
matrix is responsible for a dephasing time of the excitonic dipole
which may be much faster than the radiative recombination of the
exciton~\cite{kammerer02}. Furthermore, fluctuating charges in the
quantum dot environment can also induce significant dephasing under
non-resonant optical excitation~\cite{berthelot}. We have taken this
effect into account by introducing the parameter $\gamma^*$ in
equations~(\ref{leaks}) and we have shown that it could lead to a
drastic reduction of the contrast of the dipole induced reflection
signal. However, excitonic dephasing times limited by radiative
recombination have already been observed for a resonant excitation
of the fundamental optical transition of InAs quantum dots at low
temperature~\cite{langbein04}. Experiments aiming at the
demonstration of the giant non-linearity will in fact be performed
under similar conditions. This justifies taking $\gamma^*=0$ in the
non-linear study including leaks.

To conclude this part, let us stress the fact that quantum Rabi
oscillations~\cite{rabi} have been observed by resonantly pumping a
single quantum dot of InAs at low temperature. This observation, as
well as the successful demonstration of the coherent control of the
excitonic transition~\cite{controlecoh}, show that quantum dots can
be considered as two-level systems and used to realize
atomic-physics like experiments, provided these are properly
implemented.

\subsection{Optimization of the cavity}
We aim at optimizing the parameters of a micropillar in order to
have a maximally contrasted signal. We can experimentally control
two parameters: the intrinsic quality factor $Q_0$ and the diameter
$d$ of the micropillar. $Q_0$ corresponds to the quality factor of
the planar cavity and is tunable by changing the reflectivity of
each Bragg mirror. The diameter $d$ is adjusted during the
lithography and etching step. The total quality factor $Q$ of the
micropillar reads

\begin{equation}
\frac{1}{Q}=\frac{1}{Q_0}+\frac{1}{Q_{leak}} \; ,
\end{equation}

\noindent where the leaks are mainly due to the etching step and can
be written~\cite{rivera99}

\begin{equation}
\frac{1}{Q_{leak}}=\frac{2 |E(d)|^2 \varepsilon}{d}.
\end{equation}

\noindent $|E(d)|$ is the electrical field of the fundamental mode
at the sidewalls of the micropillar, whose profile is given by the
Bessel function of the first kind $J_0$~\cite{rivera99}. The
parameter $\varepsilon $ is a parameter quantifying the etching
quality. The leaks increase as the diameter of the etched
micropillar decrases. In the following we will take $\varepsilon
\sim 0.007$ which corresponds to realistic experimental
parameters~\cite{jmg}.

The experimental signal to maximize is defined as ${\cal
C}=T_{max}-T_{min}$, where $T_{max}$ and $T_{min}$ are given by
equations~(\ref{Tmax}) and (\ref{Tmin}), $T_{max}=(Q/Q_0)^2$ and
$T_{min}=(Q/Q_0)^2(1/(1+f))^2$ . We have chosen to optimize an
amplitude rather than a visibility ${\cal
V}=(T_{max}-T_{min})/(T_{max}+T_{min})$ because we should be then
less sensitive to the optical background. A first strategy to
optimize the contrast ${\cal C}$ is to reach small $T_{min}$, that
is high Purcell factor $F_p$, whose expression is~\cite{gerard99}

\begin{equation}
F_p=\frac{3Q}{4\pi^2V}\left(\frac{\lambda}{n}\right)^3  \; .
\end{equation}

\noindent The quantity $\lambda$ is the dipole wavelength in the
vacuum, $n$ the refractive index of the medium and $V$ the effective
volume of the mode,

\begin{equation}
V\sim \left(\frac{\lambda}{n}\right)\frac{\pi d^2}{8}.
\end{equation}

\begin{figure}[h,t]
\begin{center}
\includegraphics[height=8cm]{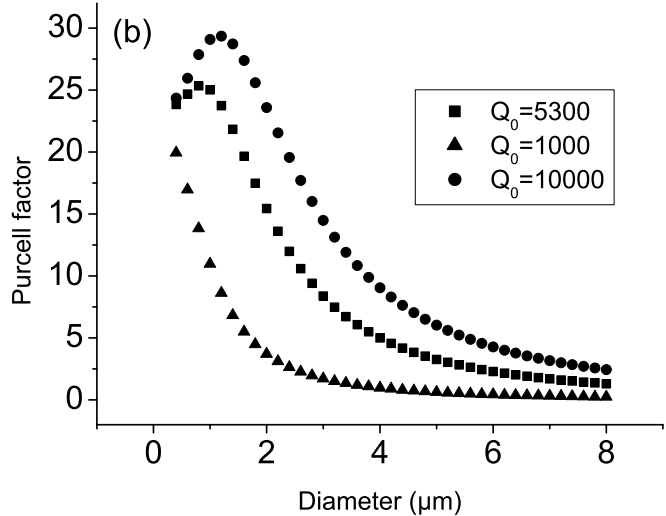}
\caption{\it Quality factor of a micropillar cavity $(a)$ and
Purcell factor of a single quantum dot in the cavity mode $(b)$ as a
function of the diameter of the micropillar, for different $Q_0$
factors of the intrinsic cavity. Squares : $Q_0=5300$. Dots :
$Q_0=10000$. Triangles : $Q_0=1000$. We took $\displaystyle{
\frac{1}{Q}=\frac{1}{Q_0}+\frac{1}{Q_{leak}}}$ with
$\displaystyle{\frac{1}{Q_{leak}}=\frac{2 |E(d)|^2 \varepsilon}{d}}
$. The quantity $|E(d)|$ is the electrical field at the sidewalls of
the micropillar.  The parameter $\varepsilon$ quantifies the leaks
due to the etching. We took here $\varepsilon=0.007$.
}\label{fig:purcell}
\end{center}
\end{figure}

\noindent Figure~\ref{fig:purcell} represents the evolution of $Q$
and $F_p$ as functions of the micropillar diameter for three
different values of the intrinsic quality factor $Q_0$ : $1000,5300$
and $10000$. If the diameter is too small, the leaks degrade $Q$ and
as a consequence $F_p$. If the diameter is too large, $F_p$
decreases because of the large modal volume. The diameters
maximizing the Purcell factor vary between $1$ and $2$~$\mu m$. As
it can be seen on the figure, a higher initial $Q_0$ allows to reach
higher values of $F_p$, and corresponds to higher optimal diameters.

\begin{figure}[h,t]
\begin{center}
\includegraphics[height=8cm]{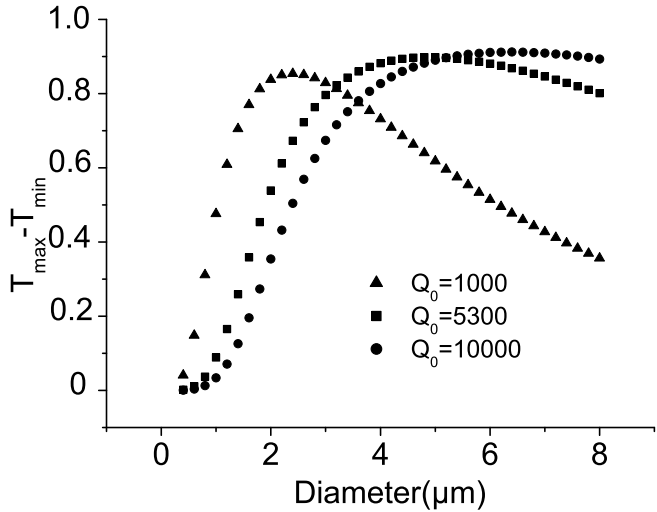}
\caption{\it Amplitude of the signal $T_{max}-T_{min}$. Squares :
$Q_0=5300$. Dots : $Q_0=10000$. Triangles: $Q_0=1000$. Amplitudes as
high as $0.9$ can be obtained using state of the art
microcavities.}\label{fig:visibility}
\end{center}
\end{figure}

At the same time we need high $T_{max}$, which corresponds to small
cavity leaks and to large diameters. We have represented in
figure~\ref{fig:visibility} the evolution of $T_{max}-T_{min}$ as a
function of the micropillar diameter for different $Q_0$. As
expected, the optimal diameters are higher than the ones obtained by
optimization of Purcell factor, and vary now between $2$ and
$6$~$\mu m$. For each $Q_0$, the amplitude of the optimized signal
is higher than $0.8$ which is quite convenient. We shall prefer the
set of parameters corresponding to the smallest diameter, so that it
is easier to isolate a single quantum dot, that is $Q_0=1000$,
$d=2.4$~$ \mu m$, $Q=960$ and $F_p=2.6$. The expected amplitude of
the experimental signal should then be $0.85$.

We shall mention here another strategy to enhance $f$, that consists
in reducing the leaks $\gamma_{at}$. Recent experiments involving
the metallization of the micropillars have shown a reduction of
$\gamma_{at}$ by a factor 10~\cite{bayer01}. The expected $T_{min}$
obtained with such a metallized cavity should be under $10^{-3}$,
and the signal amplitude near $0.9$.

\begin{figure}[h,t]
\begin{center}
\includegraphics[height=6cm]{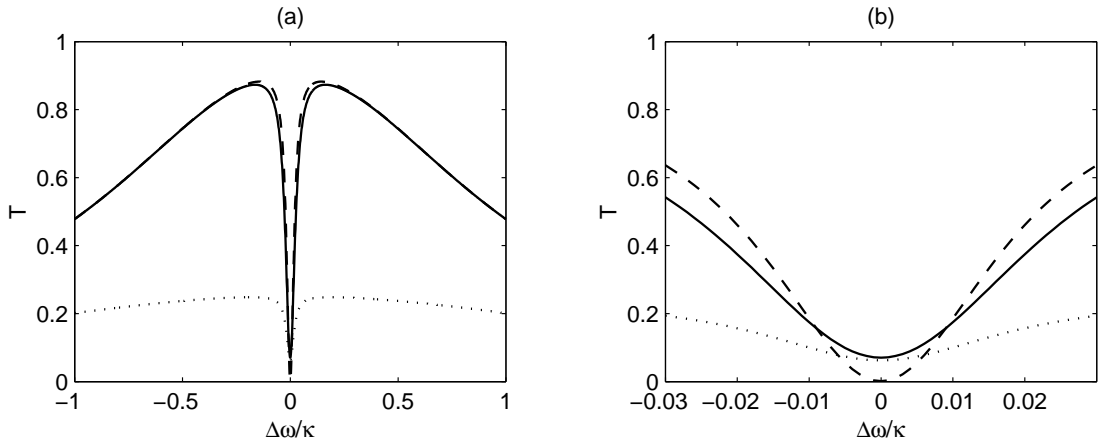}
\caption{\it $(a)$ Transmission of the optical system as a function
of the normalized detuning $\Delta \omega/ \kappa$ between the
quantum dot and the driving frequency. We took $\delta = 0$ for
convenience. Dots : we took $Q_0=1000$, $Q=500$, $F_p=3$. Solid :
$Q_0=1000$, $Q=960$, $F_p=2.6$, $d=2.4$~$\mu m$ which are the
parameters resulting from the optimization of the micropillar.
Dashed : Same parameters after metallization of the sidewalls of the
micropillar. $(b)$ Zoom on the dips. }\label{fig:transleak}
\end{center}
\end{figure}

To have a glimpse of the expected signal we have plotted in
figure~\ref{fig:transleak} the transmission of the system as a
function of the detuning between the atom and the field for
$Q_0=1000$, $Q=500$ and $F_p=3$ (dot curve) which corresponds to
realistic parameters for single photon sources before
optimization~\cite{Moreau01}. The contrast of the signal is $0.21$.
On the same figure we have also plotted the expected signal after
optimization of the micropillar, with and without metallization of
its sidewalls.

\begin{figure}[h,t]
\begin{center}
\includegraphics[height=6cm]{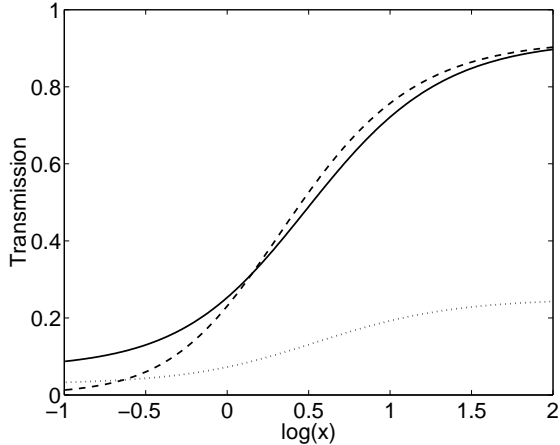}
\caption{\it Transmission of the optical system as a function of the
logarithm of the saturation parameter $\log(x)=\log (4
P_{in}/\Gamma)$. Dots : we took $Q_0=1000$, $Q=500$, $F_p=3$. Solid
: $Q_0=1000$, $Q=960$, $F_p=2.7$, $d=2.4$~$\mu m$ which are the
parameters resulting from the optimization of the micropillar.
Dashed : Same parameters after metallization of the sidewalls of the
micropillar.}\label{fig:nlresleak}
\end{center}
\end{figure}

We have finally plotted in figure~\ref{fig:nlresleak} the
transmission coefficient on resonance as a function of the logarithm
of the saturation parameter $\log(x)$ for these three different sets
of parameters. We shall be able to observe the non-linear
transmission jump with state-of-the-art~micropillar~cavities.

\subsection{Single-photon source \emph{versus} giant non-linear medium}

We have seen that the "one-dimensional atom" case requires
$f\rightarrow \infty$ and $Q/Q_0 \rightarrow 1$. Such an optical
system would also provide a high efficiency single-photon source.
The expression of the raw quantum efficiency $\eta$~\cite{barnes} of
theses devices is indeed

\begin{equation}
\eta=\frac{f}{1+f}\frac{Q}{Q_0} \;.
\end{equation}

\begin{figure}[h,t]
\begin{center}
\includegraphics[height=7cm]{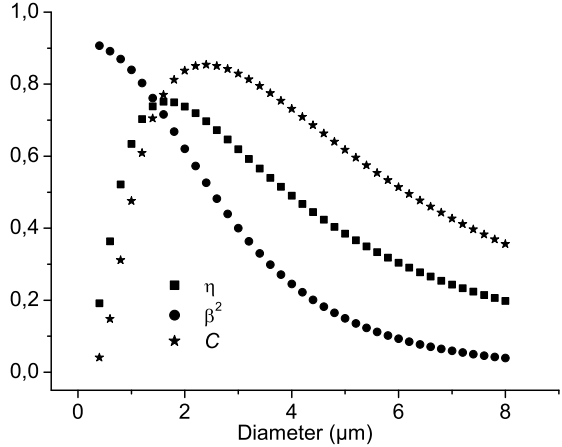}
\caption{\it Caracteristics of the quantum dot-cavity system as a
function of the diameter of the micropillar. We took $Q_0=1000$.
Squares: raw quantum efficiency $\eta$. Stars: expected amplitude
${\cal C}$ of the experimental signal. Dots : probability of photon
absorption $\beta^2$. The contrast ${\cal C}$ is more sensitive than
$\eta$ to the leaks of the cavity, leading to higher optimal
diameters. }\label{fig:beta1000}
\end{center}
\end{figure}

\noindent The prefactor $\beta=f/(1+f)$ has been introduced in
section IV, it represents the fraction of photons spontaneously
emitted by the excited atom into the cavity mode, whereas $Q/Q_0$ is
the fraction of photons initially in the cavity mode finally
funneled into the mode(s) of interest. Note that usually for
single-photon sources, the photons are collected in only one output
mode. In the present case, transmitted \emph{and} reflected photons
must be collected to measure the quantum efficiency of the
corresponding source. One may ask if optimizing the system as a
single photon source is equivalent to optimizing it as a medium
providing dipole induced transparency (one should then maximize the
visibility ${\cal C}$ of the signal) or as a giant non-linear medium
(this would require a low critical power, that is a high absorption
probability $\beta^2$ as it has been introduced in section IV). We
have plotted in figure~\ref{fig:beta1000} the parameters ${\cal C}$,
$\beta^2$ and $\eta$ as functions of the diameter of the micropillar
for an initial quality factor $Q_0=1000$. As it can be seen on the
figure, optimal diameters are different. The optimization of ${\cal
C}$ leads to the highest diameter. As it is explained in paragraph
B, this is because non-leaky cavities (and as a consequence high
diameters) are needed to reach high $T_{max}$. It is striking to
observe that $\beta^2$ and $\eta$ have different evolutions. Indeed
one could have thought that a good single-photon source, that is an
optical system that emits photons with high efficiency in a
particular mode, is also able, when it is driven by a resonant
field, to absorb and reemit photons with high efficiency. Yet
$\beta^2$ is optimized for smaller diameters that $\eta$ : the
absorption probability is more sensitive to the atomic leaks than
the single photon source efficiency. Even if the cavity is leaky, an
atom perfectly connected to the cavity mode can absorb one photon in
the input field with a maximal probability. The major difference
between these two behaviors is that they are observed in two quite
different regimes. The quantity $\eta$ is the probability of
detecting a photon in the mode of interest {\it conditioned on the
excitation of the atom}, and can be computed by supposing that in a
first step, the atom has emitted a photon in the cavity mode, and
that in a second step, this photon has been funneled into the mode
of interest. On the contrary, $\beta^2$ is estimated in a permanent
regime where the driving field can interfere with the fluorescence
field as it was pointed out in section III. A signature of this
effect has been observed in section IV, where total reflection was
induced by an atom perfectly connected to a leaky cavity. Because of
this interference phenomenon, it is impossible to describe the
evolution of the photon by successive interactions with the cavity
mode and with the atom : the atom-cavity coupled system must be
considered as a whole.

\section{Perspectives}

There has been a considerable number of proposals, e.g.
\cite{cirac,pellizzari,duan}, and experiments, e.g.
\cite{rausch,kuhn}, concerning the use of single emitters in
high-finesse cavities for quantum information processing. Most of
these papers are based on achieving the strong coupling regime. A
recent proposal relying on the Purcell regime \cite{waksprl}
requires the coherent control of additional levels in the emitter.
It is natural to ask whether the most basic non-linearity
considered in the present paper could be used directly for quantum
information applications, for example for implementing a
controlled phase gate between two photons, as suggested in Refs.
\cite{turchette95A,hofmann03}. Unfortunately recent results
suggest that this may not be possible. A numerical study
\cite{kojima04} found fidelities of quantum gates employing the
present non-linearity of order 80 \%, which is quite far from what
would be desirable for quantum computing or even quantum
communication. Higher fidelities are elusive because the
interaction with the single two-level system introduces temporal
correlations between the two input photons. An analogous
difficulty is discussed in detail in a recent theoretical paper on
the use of Kerr non-linearities for quantum computing
\cite{shapiro}. From a quantum information perspective the
relatively simple situation considered in the present paper may
thus best be seen as an important step towards the realization of
more complex configurations.

Another perspective opened by the implementation of this device
concerns the photonic computation at low threshold. As it is shown
in appendices D and E, the non-linearity studied in this paper is
not intense enough to provide bistability, but could be used to
reshape low intensity signals which may propagate in a photonic
computer. The expected performances of the device are orders of
magnitude higher than for usual saturable absorbers. Besides, if it
is fed with single photons rather than with classical fields, this
device could be operated as an all-optical switch at the single
photon level, which is a fundamental component of a photonic
computer. The theory developed in the frame of this paper could be
adapted to model such a gate and optimize its performances. This
work is under progress.

\section{Conclusion}

We have shown that a single two-level system in Purcell regime is
a medium with appealing non-linear optical properties. In the
linear case the two-level system prevents light from entering the
cavity : this is dipole induced reflectance. This property
vanishes as soon as the two-level system is saturated, which
happens for very low power, of the order of one photon per
lifetime (typically $1$nW). As a consequence, such a medium shows
a sensitivity at the single-photon level. We have established the
optical Bloch equations describing this behavior in the
semi-classical context, and shown that signatures of the
non-linearity should be observable using quantum dots and
state-of-the-art semiconducting micropillars as two-level systems
and cavities respectively. We have explored possible applications
of the non-linearity in the context of photonic information
processing.

\section{Acknowledgments}
This work is supported by the Agence Nationale de la Recherche under
the project IQ-Nona. Alexia Auffeves-Garnier is very grateful to
Xavier Letartre for the numerous and fruitful conversations.
Christoph Simon thanks Nicolas Gisin for the final reference.

\appendix

\section{Derivation of equation (\ref{diffu})}
We show in this section that $(b_r,b_t)$ and $(b_{in},b_{in}^{'})$
are related by a unitary transformation. For sake of completeness we
keep the general form for $\theta_1$ and $\theta_2$.
Equations~(\ref{q-bloch}) can be written in the stationary linear
case

\begin{equation}
S_-=\sqrt{\frac{2}{\Gamma}}\frac{ib_{in}+ib_{in}^{'}}{1+{\displaystyle
\frac{2i \Delta\omega}{\Gamma t_0(\Delta\omega)} }}  \;.
\end{equation}

\noindent As a consequence, $b_t$ reads

\begin{equation}
b_t=t_0(\Delta\omega)\left(-1+\frac{1}{1+{\displaystyle
\frac{2i\Delta\omega}{\Gamma t_0(\Delta\omega) }}}\right)b_{in}+
\left(1-t_0(\Delta\omega)+\frac{t_0(\Delta\omega)}{1+{\displaystyle
\frac{2i \Delta\omega}{\Gamma t_0(\Delta\omega)}
}}\right)b_{in}^{'} \;.
\end{equation}

\noindent It can be rewritten in the following way

\begin{equation}
b_t=-\frac{1}{1+i\zeta}b_{in}+\frac{i\zeta}{1+i\zeta}b_{in}^{'} \;
,
\end{equation}

\noindent with

\begin{equation}
\zeta=\frac{\Delta \omega +
\delta}{\kappa}-\frac{\Gamma}{2\Delta\omega}  \;.
\end{equation}

\noindent We easily compute $b_r$ by switching $b_{in}$ and
$b_{in}^{'}$. We finally obtain

\begin{equation}
\label{diffusion} \left(
\begin{array}{c}
b_r \\
b_t
\end{array}
\right)= \left(
\begin{array}{c c}
{\displaystyle \frac{i\zeta}{1+i\zeta}} & {\displaystyle \frac{-1}{1+i\zeta}} \\
{\displaystyle \frac{-1}{1+i\zeta}} & {\displaystyle
\frac{i\zeta}{1+i\zeta}}
\end{array}
\right) \left(
\begin{array}{c}
b_{in} \\
b_{in}^{'}
\end{array}
\right)  \;.
\end{equation}

\noindent The scattering matrix can be written in the following form
\begin{equation}
S= \frac{e^{i\phi}}{{\sqrt{1+\zeta^2}}}\left(
\begin{array}{c c}
\zeta & i \\
i & \zeta
\end{array}
\right)  \; ,
\end{equation}

\noindent with

\begin{equation}
\phi=\arctan \left( \frac{1}{\zeta}\right)  \;.
\end{equation}

\noindent The $S$ matrix is a unitary transformation up to a global
phase. As a consequence energy is conserved by this transformation.
Keeping in mind this property we shall rather use the
form~(\ref{diffusion}) whose coefficients have a more direct
physical interpretation.

\section{Derivation of the critical intensity including leaks}
In this section we derive the expression for the critical intensity
in the non-resonant case in presence of leaks. We use the notations
introduced in section IV. To recover the results exploited in
section III we shall impose $Q=Q_0$ and $1/f \rightarrow 0$. As it
is justified in section V, we suppose $\gamma^*=0$. The stationary
cavity population writes

\begin{equation}
a={\displaystyle t^{'}_{0}\frac{Q}{Q_0} \frac{-\Omega
S_-+\sqrt{\kappa}(ib_{in}+b_{in}^{'})+H}{\kappa }}  \; ,
\end{equation}

\noindent where $t^{'}_{0}$ has the following expression

\begin{equation}
t^{'}_{0}=\frac{1}{{\displaystyle 1 + i\frac{Q}{Q_0}\frac{\Delta
\omega + \delta }{\kappa}}}  \;.
\end{equation}

\noindent The semi-classical equations describing the evolution of
$s_z$ and $s$ write

\begin{equation}
\begin{array}{l}
\dot{s}=- i\Delta \omega s -{\displaystyle
\frac{\Gamma}{2}\frac{Q}{Q_0}\left[t^{'}_{0}+\frac{1}{f}\right] s -
i \frac{Q}{Q_0}\sqrt{\frac{\Gamma}{2}}(2s_z)b_{in}t^{'}_{0}}\\
\dot{s}_z={\displaystyle -\Gamma
\frac{Q}{Q_0}\left[\Re\left(t^{'}_{0}\right)+\frac{1}{f}\right]\left(s_z+\frac{1}{2}\right)
+\sqrt{\frac{\Gamma}{2}}\frac{Q}{Q_0}\left[is^*b_{in}t^{'}_{0}+cc\right]
}
  \;.
\end{array}
\end{equation}

\noindent where $f$, $Q$ and $Q_0$ have been defined in section III.
By sake of completeness we also give the expressions for $b_t$ and
$b_r$ after adiabatic elimination of the cavity mode

\begin{equation}
\begin{array}{l}
b_t={\displaystyle - \frac{Q}{Q_0} t^{'}_{0} b_{in} -
i \frac{Q}{Q_0} \sqrt{\frac{\Gamma}{2}}t^{'}_{0}s }
\\
b_r={\displaystyle (1- \frac{Q}{Q_0} t^{'}_{0}) b_{in} -
i\frac{Q}{Q_0} \sqrt{\frac{\Gamma}{2}}t^{'}_{0}s }
 \;.
\end{array}
\end{equation}

We obtain the stationary solution for $s$

\begin{equation}
s=-{\displaystyle i
\sqrt{\frac{2}{\Gamma}}\frac{2s_z b_{in}t^{'}_{0}}{{\displaystyle t^{'}_{0}+\frac{1}{f}+
\frac{2i\Delta \omega}{\Gamma}\frac{Q_0}{Q}}}}   \;.
\end{equation}

\noindent Injecting this solution in the evolution equation for
$s_z$, we find

\begin{equation}
\dot{s}_z=0={\displaystyle
\frac{1}{2}+s_z\left(1+\frac{|b_{in}|^2}{P_c^{'}}\right)} \; ,
\end{equation}

\noindent with

\begin{equation}
\frac{1}{P_c^{'}}=\frac{2|t^{'}_{0}|^2}{\Gamma
(\Re(t^{'}_{0})+1/f)}\left(\frac{1}{1/f+2i\frac{Q_0}{Q}\frac{\Delta\omega}{\Gamma}+t_0^{'}}+cc\right)
\; .
\end{equation}

\noindent Noting that
$2\Re(t^{'}_{0})+1/f=2/f-t^{'}_{0}-{t^{'}_{0}}^*$ we have

\begin{equation}
P_c^{'}=\frac{\Gamma}{4 |t^{'}_{0}|^2} \left(
\frac{1}{f^2}
+\frac{1}{{f}}(t^{'}_0+{t^{'}_0}^*)+\frac{2i\Delta\omega}{\Gamma}\frac{Q_0}{Q}(-t^{'}_0+{t^{'}_0}^*)+
\left(\frac{Q_0}{Q}\frac{2\Delta\omega}{\Gamma}\right)^2+ |t^{'}_0|^2\right)
\end{equation}

\noindent Let us remind here of the following expressions

\begin{equation}
\begin{array}{l}
{\displaystyle \frac{1}{|t^{'}_0|^2}=\left(\frac{Q}{Q_0}\right)^2+\left(\frac{\Delta\omega+\delta}
{\kappa}\right)^2}\\
{\displaystyle \frac{t^{'}_0+{t^{'}_0}^*}{|t^{'}_0|^2}=2}\\
{\displaystyle
\frac{-t^{'}_0+{t^{'}_0}^*}{|t^{'}_0|^2}=\frac{2iQ}{Q_0}\frac{\Delta
\omega+\delta}{\kappa}}
\end{array}
\end{equation}

\noindent We finally obtain

\begin{equation}
P_c^{'}=\frac{\Gamma}{4}\phi^{'}(\Delta \omega) \; ,
\end{equation}

\noindent with

\begin{equation}
\phi^{'}(\omega)={\displaystyle
\left(1+\frac{1}{f}\right)^2+ \left(\frac{Q}{Q_0f}\frac{\Delta \omega + \delta}{\kappa}\right)^2
+\left(\frac{2\Delta \omega}{\Gamma}\frac{Q_0}{Q}\right)^2+
\left(\frac{2\Delta \omega}{\Gamma}\frac{\Delta \omega + \delta}{\kappa}\right)^2-
\left(\frac{4\Delta \omega}{\Gamma}\frac{\Delta \omega + \delta}{\kappa}\right)} \;.
\end{equation}

\noindent If the system has no leaks, we have

\begin{equation}
\phi^{'}(\Delta \omega)=\phi(\Delta \omega)=\left(\frac{2\Delta \omega}{\Gamma}\right)^2+
\left(\frac{2\Delta \omega}{\Gamma} \frac{\Delta \omega + \delta}{\kappa}-1\right)^2
\end{equation}

\noindent Whatever the driving frequency may be, the absorption
cross section remains positive. This expression in mainly used in
section III. At resonance, we find
\begin{equation}
\phi^{'}(0)=\frac{1}{\xi}={\displaystyle
\left(1+\frac{1}{f}\right)^2} \; ,
\end{equation}
which was also exploited in section III.

\section{Slow light}

We have evidenced in section III and IV that a one-dimensional atom
is a highly dispersive medium. In particular, a quantum-dot cavity
system evanescently coupled to a waveguide has a behavior similar to
a medium showing dipole induced transparency. As a consequence, this
optical system could be used to slow down photons. Let us consider
the case of a cavity perfectly connected to a waveguide ($Q/Q_0=1$)
containing a leaky quantum dot. The transmission coefficient in
amplitude can be written $t=|t|e^{-i\phi_t(\omega)}$, where
$\phi_t(\Delta\omega)$ varies near $\omega_0$ on a scale $\Gamma$.
We send in the optical system a wave packet $\psi_{in}(\omega)$ of
width $W$ centered around $\omega_0$. Denoting $\theta$ the temporal
coordinate, we obtain the shape of the output pulse

\begin{equation}
\psi_{out}(\theta)\propto \int d\omega \psi_{in}(\omega)e^{i\omega
\theta }e^{i\phi_t(\omega)} \;.
\end{equation}

\noindent If the width of the wave packet fulfills $W \ll \Gamma$,
we can develop $\phi_t$ around $\omega_0$. We finally obtain
$\psi_{out}(\theta)=\psi_{in}\left[\theta-{\displaystyle \left(
\frac {\partial \phi_t}{\partial \omega}
\right)_{\omega_0}}\right]$. The wave packet will then be
transmitted by the optical system after a delay $T_D$ which reads

\begin{equation}
T_D=\left(\frac{\partial \phi_t}{\partial \omega
}\right)_{\omega_0} \;.
\end{equation}

\noindent During the transmission the wave packet will also be
damped by a factor $T=|t|^2$. Remembering that $\kappa \gg
\Gamma$, we neglect the variations due to the cavity mode. We
shall then take $t^{'}_{0}(\Delta\omega)\sim 1$ and

\begin{equation}
t\sim \frac{f}{1+f}\frac{1}{{\displaystyle 1+
\frac{2i}{\Gamma}\Delta \omega \frac{Q}{Q_0}\frac{f}{1+f}}} \;.
\end{equation}

\noindent With this hypothesis, $\phi_t \sim \arctan \left(
\frac{f}{1+f}\frac{2 \Delta \omega}{\Gamma}\right)$. As a
consequence,

\begin{equation}
T_D\sim \frac{2}{\Gamma} \arctan ^{'} (x)_0 \sim
\frac{2}{\Gamma}\frac{f}{1+f} \;.
\end{equation}

\noindent The wave packet is delayed by the lifetime of the
dipole, which was expected. The damping factor has the following
form

\begin{equation}
T=\left(\frac{f}{1+f}\right)^2
\end{equation}

\noindent This process could be repeated using a series of $N$
optical devices. We note $N_{1/2}$ the number of devices such that
the outcoming power is half the incoming one. $N_{1/2}$ checks

\begin{equation}
N_{1/2}=\frac{1}{2}\frac{\log{2}}{\log(1+1/f)}
\end{equation}

\noindent Supposing $f$ sufficiently high, we have
$\log(1+1/f)\sim 1/f$ and $N_{1/2}$ scales like $f$. We could
finally obtain a delay ${\cal T}_D$

\begin{equation}
{\cal T}_D=N_{1/2} \frac{2}{\Gamma}\frac{f}{1+f}\propto
\frac{\Gamma}{2} f
\end{equation}

\noindent In particular, we could use a series of microdisks each
evanescently coupled to the same wave guide. This generalizes the
study of Heebner et al~\cite{Heebner02} who have shown that the
group velocity of a signal passing through a series of empty
microdisks scales like the inverse of the finesse of the
resonators.

\section{About optical switches}

\begin{figure}[h,t]
\begin{center}
\includegraphics[height=5cm]{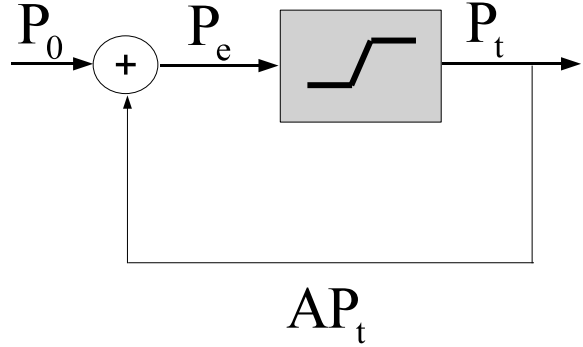}
\caption{\it Scheme of a possible use of the optical system to
generate bistability. Part of the transmitted power in reinjected at
the entrance of the device. The weak slope of the function
$T(P_{in})$ does not allow to reach the bistability regime.
}\label{fig : bistable}
\end{center}
\end{figure}

Looking at the transmission coefficient, we could be tempted to
use the giant non-linearity to realize an all-optical switch.
Bistability regime is expected to be quite useful with this
aim~\cite{notomi05,tanabe05}. As it is represented on
figure~\ref{fig : bistable}, we could for example re-inject part
of the transmitted intensity in the input port to realize a
bistable device. Unfortunately the slope of the signal is too low.
Calling $P_0$ the signal coming in the loop, $P_e$ the signal
entering the device, $P_t$ the power transmitted by the device and
$A$ the fraction of $P_t$ used to create the bistability, we have

\begin{equation}
P_e=P_0+A P_t(P_e)  \;.
\end{equation}

Bistability happens for values of the parameter $B$ for which
equation $P_0(P_e)=B$ has more than one solution. At low intensity
$P_t\sim 0$ and $P_e\sim P_0$. At high intensity $P_t\sim P_0$ and
$P_e\sim(1-A)P_0$. The system will exhibit bistability if $P_0$
decreases as $P_e$ increases. This can only be done if $\partial P_t
/ \partial P_e > 1$. Nevertheless, in can easily be shown that the
slope of the signal $P_t(P_e)$ is bounded by $(2/3)^3<1$, preventing
the system from reaching the bistability regime.

\section{Reshaping step}

\begin{figure}[h,t]
\begin{center}
\includegraphics[height=5cm]{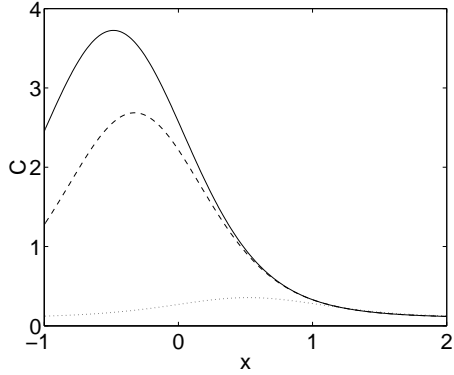}
\caption{\it Contrast enhancement ratio as a function of the
saturation parameter $x=4P_{in}/\Gamma$ for different values of
$f$. $Q_0=1000$, $Q=960$. Dot : $f=2.6$. Dashed : $f=50$. Solid :
$f=100$.}\label{fig : C}
\end{center}
\end{figure}

A possible application is to use the non-linearity to enhance the
contrast ratio between two pulses of different intensities. This
can be used to regenerate optical signals travelling in an optical
fiber. The major advantage of this system compared to other
devices is the very low switching energy, defined as the energy
necessary to saturate the system and make it switch from a linear
to a non-linear behavior. As already seen, the typical switching
energy is $0.25$~$ h \nu $ where $h\nu$ the energy of a resonant
photon. We have $h\nu \sim 1~eV \sim 4.10^{-20}J$ which is $8$
orders of magnitude lower than for traditional saturable
absorbers~\cite{oudar}. The figure of merit for this kind of
devices is the contrast enhancement ratio, defined as

\begin{equation}
{\cal
C}=\left(\frac{P_L}{P_H}\right)_{in}\left(\frac{P_H}{P_L}\right)_{t}
\end{equation}

\noindent where $P_H$ (resp $P_L$) is the high-power pulse (resp
the low one). The subscript $_{in}$ (resp~$_{t}$) describes the
incoming field (resp transmitted). We introduce the extinction
ration of the pulse

\begin{equation}
d=\left( \frac{P_H}{P_L}\right)_{in} \; .
\end{equation}

For a perfect non-linear device, ${\cal C}$ writes

\begin{equation}
{\cal C}=\frac{1}{d}\frac{T(x)}{T(x/d)} \; ,
\end{equation}

\noindent where $T$ is the transmittance at resonance of the
device. We have

\begin{equation}
{\cal C}=d \left( \frac{1+x}{1+x/d}\right)^2  \; ,
\end{equation}

\noindent  where ${\cal C}$ is maximum for $x \rightarrow 0$ and
tends to $d$. With an ideal device we could theoretically reach
any value of ${\cal C}$. Taking into account the leaks, and
denoting $T_{leak}$ the transmission of the device on resonance,
and ${\cal C}_{leak}$ the new contrast enhancement factor, we
obtain

\begin{equation} {\cal
C}_{leak}=\frac{1}{d}\frac{T_{leak}(x)}{T_{leak}(x/d)} \;.
\end{equation}

\noindent We have represented figure~\ref{fig : C} the contrast
enhancement factor for different values of the factor $f$ which has
been defined in equation~(\ref{f}). Let us recall that $f$ is
related to the Purcell factor by the simple expression
$f=(\gamma_{free}/\gamma_{at})F_p$ if there is no excitonic
dephasing. The intrinsic quality factor (resp. the quality factor)
of the cavity has been taken equal to $1000$ (resp. $950$). The
extinction ratio is doubled for $f\sim 30$, which corresponds to a
typical Purcell factor of $3$, and $\gamma_{at}/\gamma_{free}\sim
0.1$ which could be obtained by metallizing the sidewalls of a
micropillar cavity as it has been underlined in section V. The ratio
increases with $f$, a $6~dB$ enhancement is reached for $f\sim100$
which is within reach of the micropillar or photonic crystal
technology.


\begin{thebibliography}{10}

\bibitem{thompson} R.J. Thompson, G. Rempe and H.J. Kimble,
Phys. Rev. Lett. {\bf 68}, 1132 (1992).

\bibitem{brunerabi} M. Brune {\it et al.}, Phys. Rev. Lett.
{\bf 76}, 1800 (1996).

\bibitem{birnbaum} K.M. Birnbaum {\it et al.}, Nature, {\bf
436}, 87 (2005).

\bibitem{rabi} T.H. Stievater {\it et al.}, Phys. Rev. Lett. \textbf{87}, 133603
(2001); H. Kamada, H. Gotoh,J. Temmyo,T. Takagahara, H. Ando, Phys.
Rev. Lett. \textbf{87}, 246401 (2001).

\bibitem{controlecoh} T. Flissikowski, A. Betke, I. A. Akimov and F. Henneberger, Phys. Rev. Lett.
\textbf{92}, 227401 (2004); Q.Q. Wang  {\it et al.}, Phys. Rev.
Lett. \textbf{95}, 187404 (2005).

\bibitem{semicon} J.P. Reithmaier {\it et al.}, Nature {\bf
432}, 197 (2004); T. Yoshie {\it et al.}, Nature {\bf 432}, 200
(2004); E. Peter {\it et al.}, Phys. Rev. Lett. {\bf 95}, 067401
(2005).


\bibitem{purcell46} E. M. Purcell, Phys. Rev. \textbf{69}, 681
(1946).

\bibitem{gerard99} J. M. Gerard {\it et al.}, Phys. Rev. Lett. \textbf{81},
1110 (1998)

\bibitem{Solomon} G.S. Solomon, M. Pelton and Y. Yamamoto, Phys. Rev. Lett. \textbf{86}, 3903
(2001).

\bibitem{Moreau01} E. Moreau {\it et al.}, Applied Physics Letters \textbf{79}, 2865
(2001).

\bibitem{Var05} C. Santori {\it et al.}, Nature \textbf{419}, 594 (2002);
S. Varoutsis {\it et al.}, Phys. Rev. B. \textbf{72}, 041303(R)
(2005).

\bibitem{turchette95A}  Q.A. Turchette, C.J. Hood, W. Lange, H. Mabuchi and H.J. Kimble, \emph{Phys. Rev. Lett.} \textbf{75}, 4710
(1995).

\bibitem{wakspra} E. Waks and J. Vuckovic, Phys. Rev. A
{\bf 73}, 041803(R) (2006).

\bibitem{waksprl} E. Waks and J. Vuckovic
Phys. Rev. Lett. {\bf 96}, 153601 (2006).


\bibitem{hofmann03} H.F. Hofmann, K. Kojima, S. Takeuchi and K. Sasaki, Journal of Optics B, {\bf 5}, 218
(2003).

\bibitem{turchette95B} Q. A. Turchette, R. J. Thompson and H. J. Kimble, Applied Physics B \textbf{60}, S1-S10
(1995).


\bibitem{Gardiner85} C. W. Gardiner and M. J. Collett, Phys. Rev. A, {\bf 31}, 3761
(1985).

\bibitem{akahane05} Y. Akahane et al, Optics Express \textbf{13},
2512 (2005).

\bibitem{fan05} J. T. Shen and S. Fan, Optics Letters \textbf{30}, 2001
(2005).

\bibitem{kojima04} K. Kojima, H.F. Hofmann, S. Takeuchi and K. Sasaki, Phys. Rev. A \textbf{70}, 013810
(2004).

\bibitem{Allen} Allen and Eberly, \emph{Optical resonance and two-level
systems}, Dover.

\bibitem{Heebner02} J.E. Heebner, R.W. Boyd, Q-Han Park, Phys. Rev. E. \textbf{65}, 036619
(2002).


\bibitem{kuhn} A. Kuhn, M. Hennrich and G. Rempe,  Phys. Rev. Lett. {\bf 89}, 067901
(2002).

\bibitem{cohen} C. Cohen-Tannoudji {\it et al.}, \emph{Processus d'interaction
entre Photons et Atomes}, Ed. CNRS, p.366.

\bibitem{said} A. A. Said {\it et al.}, J. Opt. Soc. Am. B \textbf{9}, 405
(1992).

\bibitem{Gra98} J.P. Poizat and P. Grangier, Phys. Rev. Lett. \textbf{70}, 271
(1993); J.F. Roch et al, Phys. Rev. Lett. \textbf{78}, 634 (1997);
P. Grangier et al, Nature \textbf{396}, 537 (1998).


\bibitem{Hau99} L. V. Hau, S. E. Harris, Z. Dutton and C. W. Behroozi, Nature \textbf{397}, 594 (1999).

\bibitem{rivera99} T. Rivera {\it et al.}, Applied Physics Letters \textbf{74}, 911
(1999).

\bibitem{jmg} J. M. Gerard, Topics of Applied Physics \textbf{30}, 269
(2003).

\bibitem{jmg2} J. M. Gerard {\it et al.}, J. Cryst. Growth \textbf{150}, 351
(1995).

\bibitem{marzin} J.Y. Marzin, J.M. Gerard, A. Izrael, D. Barrier,G. Bastard, Phys. Rev. Lett. \textbf{73}, 716
(1994).

\bibitem{karrai} S. Seidl {\it et al.}, Phys. Rev. B \textbf{72},
195339 (2005).

\bibitem{bayer} M. Bayer {\it et al.}, Phys. Rev. B \textbf{65}, 195315
(2002).

\bibitem{paillard} M. Paillard {\it et al.}, Phys. Rev. Lett. \textbf{86},
1634 (2000).

\bibitem{kammerer02} C. Kammerer {\it et al.}, Phys. Rev. B \textbf{66}, 041306(R) (2002).

\bibitem{berthelot} A. Berthelot {\it et al.}, Nature Physics \textbf{2},
759 (2006).

\bibitem{langbein04} W. Langbein {\it et al.}, Phys. Rev. B \textbf{70},
033301 (2004).

\bibitem{bayer01} M. Bayer {\it et al.}, Phys. Rev. Lett. \textbf{86}, 3168
(2001).

\bibitem{barnes} W. L. Barnes {\it et al.}, Eur. Phys. J. D \textbf{18}, 197
(2002).

\bibitem{notomi05} M. Notomi \emph{et al.}, Optics Express \textbf{13}, 2678 (2005);

\bibitem{tanabe05} T. Tanabe \emph{et al.} , Optics Letters \textbf{30}, 2575 (2005);

\bibitem{oudar} J. Mangeney {\it et al.}, Electronic Letters \textbf{36}, 1486
(2000).


\bibitem{cirac} J.I. Cirac, P. Zoller, H. J. Kimble and H. Mabuchi, Phys. Rev. Lett.
{\bf 78}, 3221 (1997).

\bibitem{pellizzari} T. Pellizzari, S. A. Gardiner, J. I. Cirac and P. Zoller, Phys. Rev.
Lett. {\bf 75}, 3788 (1995).

\bibitem{duan}  L.-M. Duan and H. J. Kimble,
Phys. Rev. Lett. \textbf{92}, 127902 (2004).

\bibitem{rausch} A. Rauschenbeutel {\it et al.}, Phys. Rev.
Lett. {\bf 83}, 5166 (1999).

\bibitem{shapiro} J.H. Shapiro, Phys. Rev. A {\bf 73},
062305 (2006).








\end{thebibliography}
\end{document}